\documentclass[preprint,showpacs,preprintnumbers,amsmath,amssymb]{revtex4}

\usepackage{graphicx}
\usepackage{dcolumn}
\usepackage{bm}

\newcommand{\be}{\begin{equation}}
\newcommand{\ee}{\end{equation}\noindent}
\newcommand{\bea}{\begin{eqnarray}}
\newcommand{\eea}{\end{eqnarray}}
\newcommand{\Tr}{\mbox{Tr}}

\newcommand{\xo}{x_1}
\newcommand{\xop}{x_{1}{\large '}}

\newcommand{\xt}{x_2}
\newcommand{\xtp}{x_{2}{\large '}}

\newcommand{\vxo}{\vec{x}_1}
\newcommand{\vxop}{\vec{x}_{1}{\large '}}
\newcommand{\vyo}{\vec{y}_1}

\newcommand{\vxt}{\vec{x}_2}
\newcommand{\vxtp}{\vec{x}_{2}{\large '}}
\newcommand{\vyt}{\vec{y}_2}

\newcommand{\Gs}{{G^{(s)}}}
\newcommand{\gf}{\gamma_5}
\newcommand{\vpn}{\vec{p}_n}
\newcommand{\vpm}{\vec{p}_m}
\newcommand{\Svxo}{\sum_{\vxo}}
\newcommand{\Svxt}{\sum_{\vxt}}
\newcommand{\Svxop}{\sum_{\vxop}}
\newcommand{\Svxtp}{\sum_{\vxtp}}
\newcommand{\Svyo}{\sum_{\vyo}}
\newcommand{\Svyt}{\sum_{\vyt}}
\newcommand{\FTpmo}{\frac{e^{+i\vpm\cdot\vxo}}{L^3}}
\newcommand{\FTpmt}{\frac{e^{-i\vpm\cdot\vxt}}{L^3}}
\newcommand{\FTpno}{\frac{e^{+i\vpn\cdot\vxop}}{L^3}}
\newcommand{\FTpnt}{\frac{e^{-i\vpn\cdot\vxtp}}{L^3}}

\newcommand{\AdjFTpmt}{\frac{e^{+i\vpm\cdot\vxt}}{L^3}}
\newcommand{\AdjFTpno}{\frac{e^{-i\vpn\cdot\vxop}}{L^3}}

\newcommand{\xxo}{\vxo,t_S}
\newcommand{\xxt}{\vxt,t_S}
\newcommand{\xxop}{\vxop,t}
\newcommand{\xxtp}{\vxtp,t}
\newcommand{\yyo}{\vyo,t_S}
\newcommand{\yyt}{\vyt,t_S}
\newcommand{\Sj}{\frac{1}{N_R}\sum_{j}\quad}
\newcommand{\Sjo}{\frac{1}{N_R}\sum_{j_1}\quad}
\newcommand{\Sjt}{\frac{1}{N_R}\sum_{j_2}\quad}

\newcommand{\la}{\langle}
\newcommand{\ra}{\rangle}

\begin{document}


\title{Lattice Study of $K\pi$ Scattering in $I$ = 3/2 and 1/2}

\author{J. Nagata}%
  \affiliation{%
  Faculty of Informatics, Hiroshima Kokusai Gakuin University, Hiroshima 739-0321, Japan
  }%
  \altaffiliation[Present address: ]{Admissions Center, Hiroshima University, Higashi-Hiroshima 739-8511, Japan}

\author{S. Muroya}
  \affiliation{%
Department of Comprehensive Management, Matsumoto University, 
Matsumoto 390-1295, Japan}

\author{A. Nakamura}
  \affiliation{
  Information Media Center, Hiroshima University, Higashi-Hiroshima 739-8521, Japan
  }%

\date{\today}

\begin{abstract}
We report the first lattice QCD results of the scattering amplitudes of the $K\pi$ system for $I = 1/2$
channel together with  $I=3/2$ case.
We investigate all quark diagrams contributing to these iso-spin states, and find that the scattering
amplitudes are expressed as combinations of only three diagrams after setting the masses of
$u$-quark and $d$-quark to be the same. 

The lattice simulations are performed in the quenched approximation at
$\beta = 2.23$ on a 12$^3 \times 24$ lattice with an improved Iwasaki gauge action. 
We employ a new dilution-type noise method to get accuracy of data  with reasonable CPU time.
A simple method is proposed and applied to eliminate lattice artefact due to the finite extent of lattice 
along the time direction.

A clear difference in the quark mass dependence between $I=3/2$ and $I=1/2$ channels
is observed.
Although the chiral extrapolation is subtle, we assume $E_{K\pi}^2 \propto m_{u,d}^2 $,
and obtain the $S$-wave scattering lengths as
$a_0(I=3/2) m_{\pi} =-0.084^{+0.051}_{-0.064}$ 
and 
$a_0(I=1/2) m_{\pi} = -0.625\pm0.012$.
We show all necessary formulas which make the calculation possible.

We argue that $\Lambda N$ is the most appropriate target of  the L\"uscher's formula
for baryonic system because it has no $\pi$ exchange diagrams and has a 
scattering length suitable for a lattice QCD simulation.

\end{abstract}

\maketitle

\section{\label{sec:INTRODUCTION}INTRODUCTION}

One of the main objectives in our study of hadron nuclear physics is to describe
hadron interactions on the basis of QCD (quantum chromodynamics).  
The study of hadron interactions will help us understand such interactions
 in terms of multi-quark reactions mediated by gluons.
%
This is still simply a dream, and we need using phenomenological models whose parameters are fitted to the experimental data.
%
However, even this strategy is still difficult, particularly when the strangeness is
included, because in such a case, there is limited experimental information available.

The development of high-energy hadron accelerators such as those at JLab, LEPS, and J-PARC and hadron studies carried out at these facilities have contributed to the accumulation of experimental data pertaining to high-energy quark interactions. In these accelerators, not only $u$ and $d$ quarks but also $s$ quarks are excited, and high-statistics studies on hadronic reactions have been carried out; numerous hypernuclei have also been produced in these reactions

The study of hadronic reactions in a
unified manner through first-principle calculation is very important,
then, quark-gluon reactions can be studied on the basis of QCD
by analyzing the large amount of precise data obtained at the aforementioned facilities.  
The results of recent lattice calculations of the $NN$ force are very
encouraging in this direction\cite{Ishii08}.
We will step into a new era of hadron physics.

L\"uscher derived the basic formula for the calculation of scattering 
lengths on the basis of lattice QCD simulations, where the $S$-wave scattering length $a_0$ 
between 
two hadrons is related to the energy shift of the two-hadron 
state
that is 
confined in a finite
periodic spatial box of size $L^3$ at zero relative momentum\cite{Lus83,Lus86a,Lus86b,Lus91}. 
Meson-meson,
meson-baryon,  and baryon-baryon scattering lengths have been studied 
recently by using L\"uscher's formula \cite{Fuk95,Aok03,Yam04,Mur04,Mia04,Bea06,Mur07}. 
In 
these 
calculations, contributions from different 
types
of quark diagrams to the hadron four-point functions have been analyzed.
L\"uscher's formula is expected to play a
major  
role in the study of  hadronic reactions
on the basis of lattice QCD.  
In principle, there are no limitations for including strangeness
and other flavor degrees of freedom, and no free parameters.

We have calculated 
scattering lengths in the $\Lambda p$ system \cite{Mur04}
using L\"uscher's formula.
This method involves a long calculation time. To reduce the calculation time, we use 
the modified noise method 
in the present study 
for the evaluation of quark propagators.

In this 
research, we study $K \pi$ scattering, which is  
a simple but important
fundamental reaction.
This reaction is interesting because of reasons as follows :
\begin{enumerate}
\item
It is the simplest reaction
that includes $s$ quark. 
\item
The $I=1/2$ channel of this reaction is directly related to the scalar
meson $\kappa$ \cite{Wada07}. Further, this reaction is easier to study
than the $\pi\pi$ scattering reaction, in which case $I=0$ is directly related to 
the scalar meson $\sigma$.
\item
The force between $K$ and $\pi$ may produce a $K \pi N$ bound state, which
can be used to explain the penta-quark state \cite{Nak03,Kishimoto03}.
\item
Direct lattice QCD measurement of $I=1/2$ and $3/2$ will provide
a test of the validity of chiral perturbation with strangeness.
\end{enumerate}

\noindent
In the present study, we have evaluated the scattering length of the $K \pi$ system 
by lattice QCD owing to the above mentioned features of $K \pi$ system.

In Section II, we explain  the formulation based on L\"uscher's formula 
for the $K \pi$ system.
In the $K \pi$ system, there are 22 quark diagrams.
The number of independent diagrams can be  
reduced to six if we assume the masses of $u$ and $d$ quarks to be identical.
After simple calculation related to the iso-spin states,  only three diagrams contribute to the $I = 1/2$ state and two to 
the $I = 3/2$ state.
In Section III,  we show   the results
obtained in our simulations for the iso-spin channels $I = 3/2$ and $1/2$. In Section IV, we discuss 
the differences in the contributions from each diagram. The final section includes concluding remarks.

\section{\label{sec:Method}METHOD}
\subsection{\label{sec:luscherA}Scattering length determined using L\"usher's formula}

L\"uscher's formula which relate the energy shift $\Delta E$ to the scattering length\cite{Lus83} 
is given as
\begin{eqnarray}
\Delta E &=& E_{K \pi} - (m_K + m_{\pi}) \cr
         &=& - { 2 \pi (m_K + m_{\pi}) a_0 \over  m_\pi m_K L^3}
           \bigl[ 1 + c_1 { a_0 \over L} + c_2 ( {a_0 \over L} )^2 \bigr] + O(L^{-6})
\label{luesher}
\end{eqnarray}
with $c_1= -2.837297$ and  $c_2= 6.375183$, where $E_{K\pi}$ is 
the total energy of $K\pi$ system, $m_K$ and $m_{\pi}$ are masses 
of $K$ and $\pi$, and  $L$ represents the spatial size of 
lattice, respectively.

Rummukainen and Gottlieb extended the  above  formula to moving frames 
\cite{Rum95} and succeeded in calculating phase shifts in addition to the scattering length.

Using operators $O_K(x_1)$ and $O_\pi(x_2)$ for $K$ and $\pi$ 
at points $x_1$ and $x_2$, respectively, we represent 
hadron four-point functions as follows :
\be
C_{K \pi}(x_1',x_2',x_1,x_2) 
  = \bigl< O_K(x_1') O_{\pi}(x_2') O_K^{\dagger}(x_1) 
  O_{\pi}^{\dagger}(x_2)\bigr>. 
\ee
Here, $\la\cdots\ra$ represents the expectation value of the path integral,
which we evaluate using quenched lattice QCD simulations. 

After obtaining the sum over spatial coordinates 
$\vec{x}_1$, $\vec{x}_2$, $\vec{x}_1'$ and $\vec{x}_2'$,
we obtain the four-point function in the zero-momentum state, 
whose behavior is given below :
\bea
\sum_{\vec{x}_1'}\sum_{\vec{x}_2'}\sum_{\vec{x}_1}\sum_{\vec{x}_2}
C_{K \pi}(x_1',x_2',x_1,x_2) 
\nonumber \\
 =
Z_{K\pi}\cosh(E_{K\pi}(t-N_t/2))&+&Z_{K\pi}'\cosh(E_{K\pi}'(t-N_t/2)) + \cdots. 
\eea
Here, $x_1'=(\vec{x}_1',t_1')$,  $x_2'=(\vec{x}_2',t_2')$, 
$x_1 = (\vec{x}_1,t_1)$, and $x_2 = (\vec{x}_2,t_2)$ with $t_1'=t_2'$ and $t_1=t_2$.
$t$ stands for the time difference, $t\equiv t_2 - t_1$. 
$E_{K\pi}$ and $E_{K\pi}'$ are the ground and excited levels, respectively.
Hadron two-point functions are also given by
\begin{eqnarray}
C_K(x_1)  
  &=& \bigl< \sum_{x_1'}O_K(x'_1) \sum_{x_1}O^{\dagger}_K(x_1) \bigr> 
\nonumber \\
  &=& Z_K\cosh(m_K(t-N_t/2)) +  Z_K'\cosh(m_K'(t-N_t/2)) + \cdots 
\nonumber \\
C_{\pi}(x_1)  
  &=& \bigl< \sum_{x_2'}O_{\pi}(x'_2) \sum_{x_1}O^{\dagger}_{\pi}(x_2) \bigr> 
\nonumber \\
  &=& Z_{\pi}\cosh(m_{\pi}(t-N_t/2)) + Z_{\pi}'\cosh(m_{\pi}'(t-N_t/2)) + \cdots .
\end{eqnarray}

The energy shift $\Delta E$ can be deduced directly as the difference between 
$E_{K\pi}$ and $E_{K}+E_{\pi}$, which are obtained from our simulations of 
the four-point function $C_{K \pi}$ and the two-point functions, $C_K$ and $C_{\pi}$.
In the present study, the final values of $\Delta E$ are obtained from 
the results of the fitting procedure for
$C_{K \pi}$, $C_{K}$ and $C_{\pi}$. 
The details will be explained in Section III.

\subsection{\label{sec:luscherB}
Quarks diagrams in $K\pi$ scattering}

\begin{figure}[htb]
\centerline{
\includegraphics[width=.8 \linewidth]{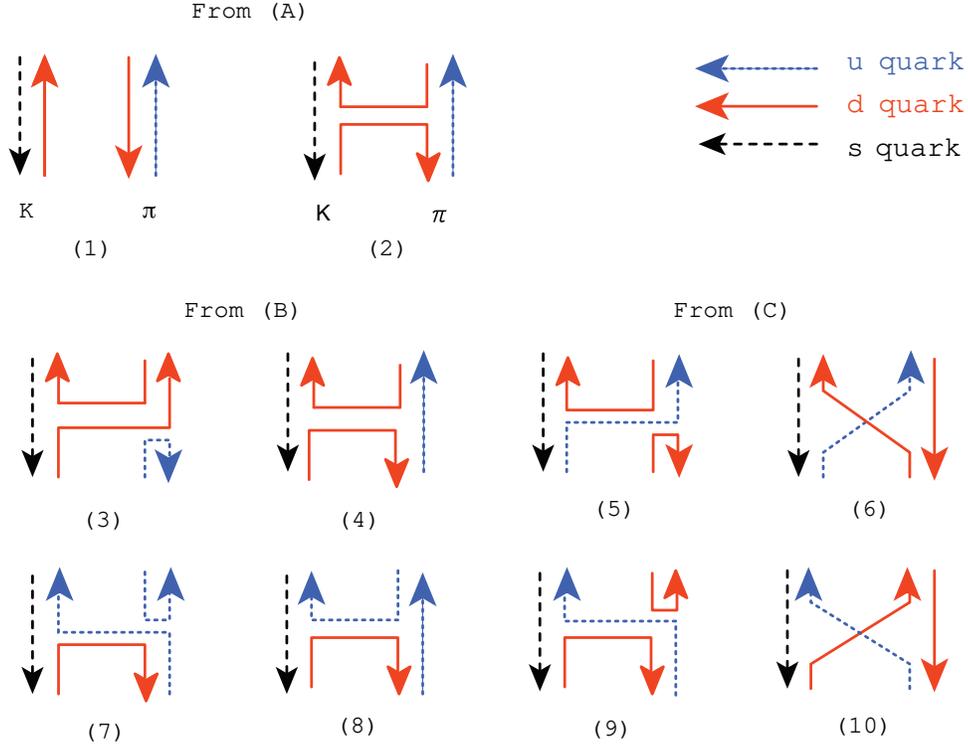}
}
\caption{
Quark propagators corresponding to Eq. $A1$-$A10$ in Appendix A.
\label{fig:QD1} 
}
\end{figure}

\begin{figure}[htb]
\centerline{
\includegraphics[width=.7 \linewidth]{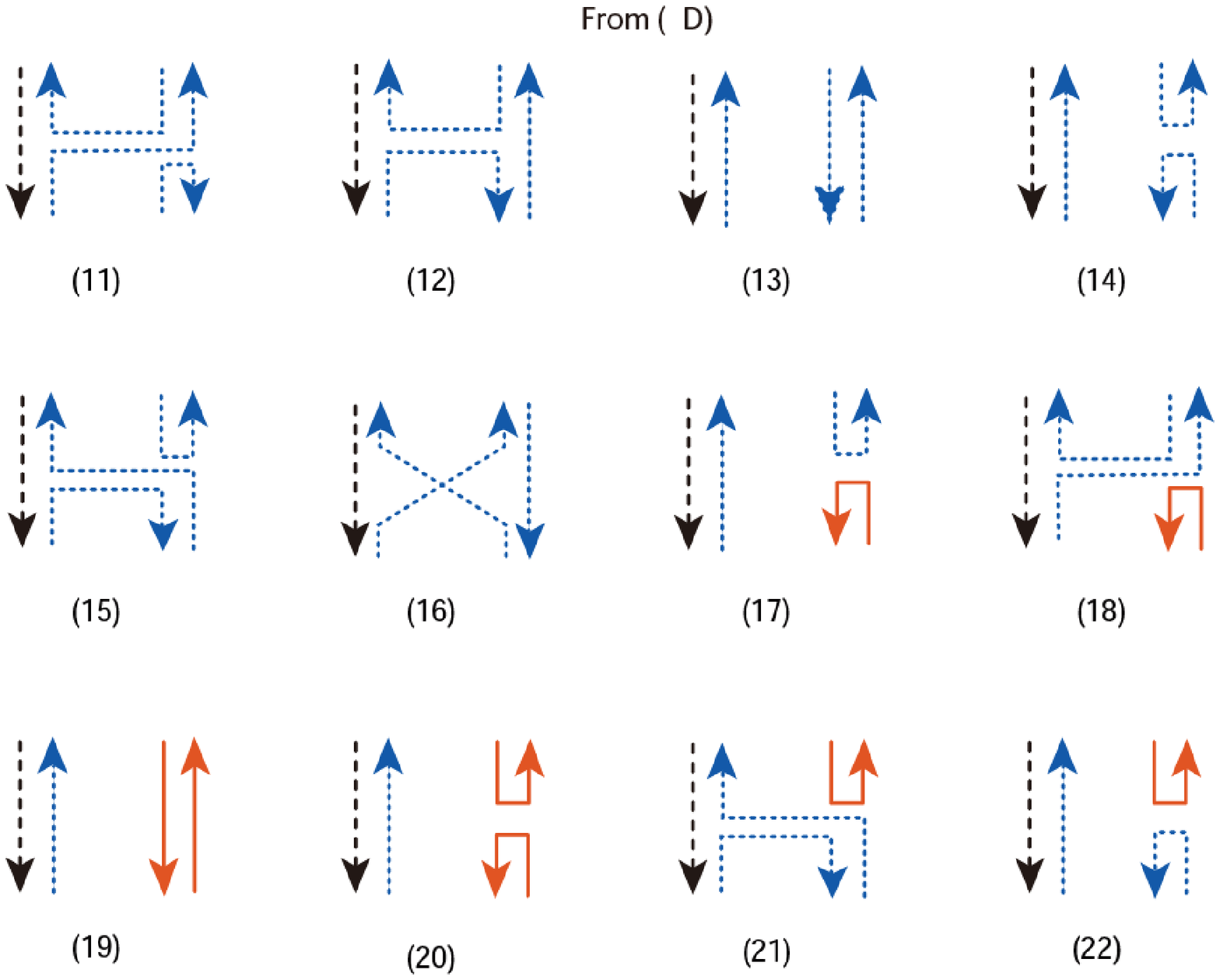}
}
\caption{
Quark propagators corresponding to Eqs. $A11$-$A22$ in Appendix A.
\label{fig:QD2} 
}
\end{figure}

In the $K \pi$ system, there are two iso-spin states, $I = 3/2$ and $I=1/2$,
\bea
|K\pi(I=1/2)>&=& \sqrt{2 \over 3}|K^0>|\pi^+>-{1 \over \sqrt{3} }|K^+>|\pi^0>
\label{eq:CG-coeff1} , \\
|K\pi(I=3/2)>&=& |K^+>|\pi^+> .
\label{eq:CG-coeff2}
\eea
Then, the scattering amplitude $M$ for $K\pi$ scattering in the $I=1/2$ state is given by
\begin{eqnarray}
M(I=1/2) &=& 
{ 2 \over 3} <K^0\pi^+|S|K^0\pi^+> - { \sqrt{2} \over 3} <K^0\pi^+|S|K^+\pi^0>
\nonumber \\
&-& { \sqrt{2} \over 3} <K^+\pi^0|S|K^0\pi^+> + { 1 \over 3} <K^+\pi^0|S|K^+\pi^0> 
\end{eqnarray} 
We introduce  the following operators explicitly for $O_K(x)$ and $O_{\pi}(x)$ :

\begin{eqnarray}
%
O_{K^0}(x) &=& \bar{s}(x)\gamma_5 d(x), \cr
O_{K^+}(x) &=& \bar{s}(x)\gamma_5 u(x), \cr
O_{\pi^+}(x)&=&-\bar{d}(x)\gamma_5 u(x), \cr
O_{\pi^0}(x)&=&{ 1 \over \sqrt{2} } \Bigl\{  \bar{u}(x)\gamma_5 u(x) - \bar{d}(x)\gamma_5 d_(x)   \Bigr\}. 
\end{eqnarray}
We insert Eq. (8) into Eq. (2) to obtain the quark diagrams for $K\pi$ scattering.
The details of our calculations are provided in Appendix \ref{sec:AppendixA}. in Eqs. A1 - A22.
Finally, we have 22 different diagrams for $I=1/2$,  
as shown in Figs.\ \ref{fig:QD1} and \ref{fig:QD2} , which correspond to
Eqs. A1 - A22. 

\begin{figure}[htb]
\centerline{
\includegraphics[width=.7 \linewidth]{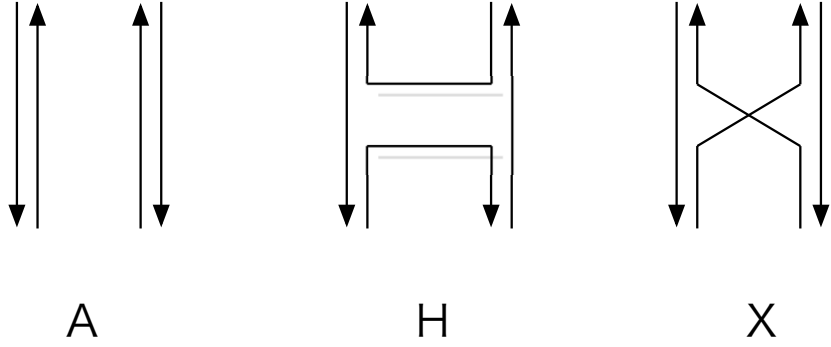}
}
\caption{
Diagrams $A$, $H$, and $X$.
\label{fig:AHX} 
}
\end{figure}

However,   by assuming that 
 $u$ and $d$ quarks have the same mass, we can categorize the 22 diagrams into six independent groups. 
Diagrams\ 1, 12, and 22 in Figs.\ \ref{fig:QD1} and \ref{fig:QD2} are compiled 
into Group 1. Similarly, diagrams\ 11, 17, 19 and 21 are compiled 
into Group 2, No.\ 2, 4, 8, and 13 in Group 3, No.\ 6, 10, and 15 in Group 4, 
diagrams\ 3, 5, 14, and 18 to Group 5, and  
diagrams\ 7, 9, 16, and 20 Group 6, respectively.
According to Eq. (8) and 
Eqs. A1 - A22, the weights of each of these groups are given as follows:

\begin{eqnarray}
\hbox{Group 1:} &&\cr
&&{2 \over 3}+{1 \over 3 } ( - { 1\over 2} ) ( -1) + { 1 \over 3} ( - { 1 \over 2} ) ( -1 ) = 1 : A, \cr
\hbox{Group 2:} &&\cr
&&({ 1 \over 3})(-{1\over 2}) + ({1\over 3})(-{1\over 2})  + ({1 \over 3})(-{1\over 2})(-1) + ( {1\over 3})(-{1\over 2}) = 0, \cr
\hbox{Group 3:} &&\cr
&&({ 2 \over 3})(-1) + (- {\sqrt{2} \over 3})(-{1\over \sqrt{2}})  + (-{\sqrt{2} \over 3})(-{1\over \sqrt{2}})(-1)  + ( {1\over 3})(-{1\over 2}) = -{3 \over 2} : H, \cr
\hbox{Group 4:} &&\cr
&&(- {\sqrt{2} \over 3})(-{1\over \sqrt{2}})(+1)  +  (- {\sqrt{2} \over 3})({1\over \sqrt{2}})(-1) + ({1 \over 3}) (-{1 \over 2}) = { 1 \over 2} : X, \cr
\hbox{Group 5:} &&\cr
&&(- {\sqrt{2} \over 3})(-{1\over \sqrt{2}})  +  (- {\sqrt{2} \over 3})({1\over \sqrt{2}})  + ({1 \over 3}) (-{1 \over 2})(-1)(-1) + ({1 \over 3}) (-{1 \over 2})(-1) = 0, \cr
\hbox{Group 6:} &&\cr
&&(- {\sqrt{2} \over 3})(-{1\over \sqrt{2}})  +  (- {\sqrt{2} \over 3})({1\over \sqrt{2}})  + ({1 \over 3}) (-{1 \over 2})(-1) + ({1 \over 3}) (-{1 \over 2})(-1)(-1) = 0.\cr
\end{eqnarray}
Here,  $A$, $H$, and $X$  are denoting the type of the quark diagrams 
shown in Fig.\ \ref{fig:AHX}.
Finally, only the three diagrams shown in Fig.\ \ref{fig:AHX} 
remain to contribute to the $K \pi$ scattering amplitudes. 
Ultimately, both $I=3/2$ and $1/2$ channels can 
be expressed by using  diagrams $A$, $H$, and $X$ as follows:
\begin{eqnarray}
M(I=3/2) &=& A - X, 
\\
M(I=1/2) &=& A - { 3 \over 2} H + { 1 \over 2} X.
\end{eqnarray}

$A$, $H$, and $X$, are schematically shown in Fig. \ref{fig:AHX}
and  are given in terms of the quark propagators
$G$ as
\bea
A(\xop,\xtp,\xo,\xt)
&=&
\Tr\left(
G(\xop,\xo) \gf \Gs(\xo,\xop) \gf \right)
\times
\Tr\left(
G(\xtp,\xt) \gf G(\xt,\xtp) \gf \right)
\nonumber \\
&=&
\Tr\left(\gf \Gs(\xo,\xop) \gf G(\xop,\xo) \right)
\times
\Tr\left(\gf G(\xt,\xtp) \gf G(\xtp,\xt) \right),
\nonumber \\
H(\xop,\xtp,\xo,\xt)
&=&
\Tr\left(
G(\xtp,\xt) \gf G(\xt,\xo) \gf 
\Gs(\xo,\xop) \gf G(\xop,\xtp) \gf \right)
\nonumber \\
&=&
\Tr\left(
\gf \Gs(\xo,\xop) \gf G(\xop,\xtp) \gf 
G(\xtp,\xt) \gf G(\xt,\xo) \right),
\nonumber \\
X(\xop,\xtp,\xo,\xt)
&=&
\Tr\left(
G(\xop,\xt) \gf G(\xt,\xtp) \gf 
G(\xtp,\xo) \gf \Gs(\xo,\xop) \gf \right)
\nonumber \\
&=&
\Tr\left(
\gf \Gs(\xo,\xop) \gf 
G(\xop,\xt) \gf G(\xt,\xtp) \gf 
G(\xtp,\xo) 
\right).
\label{eq:AHX}
\eea
Here, Tr stands for the trace over color and Dirac indices.

\subsection{\label{sec:MOMENTUM}Calculation of quark propagators using noise vectors}

Now, we calculate the four-point functions in spatial momentum space
using the Fourier transform of Eq.\ (\ref{eq:AHX}) at fixed $t$.
In a standard lattice QCD simulation, quark propagators $G (=D^{-1})$
are calculated by inverting the quark matrix $D$,
\be
D\vec{X} = \vec{B},
\label{eq:solver}
\ee
using a conjugate gradient type solver.
While calculating the hadron four-point functions, we obtain the form
\be
\sum_{\vec{x}} e^{i\vec{p}\vec{x}} \Tr \left[ D^{-1}(\vec{x},t;..) 
\cdots D^{-1}(..,\vec{x},t) \right].
\label{eq:TrDinvDinv}
\ee
For example, in the case of meson-meson scatterings 
 composed of four quark lines,
four quark propagators, $D^{-1}$, appear inside $\Tr[\cdots]$ 
in Eq.\ (\ref{eq:TrDinvDinv}).

If we calculate all necessary components of $D^{-1}$ with
Eq.\ (\ref{eq:solver}), a huge computational resource is required.
In order to reduce the simulation cost,
we introduce noise vectors 
\be
\sum_{j=1}^{N_R}
\xi_{j}(\vec{x})^{\dagger} \xi_{j}(\vec{y})
= \delta_{\vec{x},\vec{y}}
\label{eq:noise}
\ee
and rewrite Eq.\ (\ref{eq:TrDinvDinv}) as
\be
\frac{1}{N_R} \sum_{j}
\sum_{\vec{x}} e^{i\vec{p}\vec{x}} \Tr \left[ 
\xi_j(\vec{x})^{\dagger} D^{-1}(\vec{x},t;..) 
\cdots D^{-1}(..,\vec{y},t) \xi_j(\vec{y}) \right].
\label{eq:TrDinvDinv2}
\ee

Then the four-point functions in the momentum space
can be written as
\be
A, \ \ X, \ \ H = \sum \vec{Y}^{\dagger} \vec{Z},
\ee
where
\be
\vec{Y} \ \ \ \mbox{or} \ \ \ \vec{Z} = \cdots D^{-1} \cdots \xi .
\ee
Explicit formulas are given in Appendix \ref{sec:AppendixB}; 
We describe explicitly where the noise vectors are inserted.

In Eq.(\ref{eq:TrDinvDinv}), not only $\vec{x}$ but also the color and Dirac indices
are summed up because of $\Tr$.  
Then one may extend Eq.(\ref{eq:noise}) to include the color and Dirac degrees of freedom,
which will reduce CPU time further.  However, we do not take this 
in order to keep signal to noise ratio at reasonable levels.


\section{\label{sec:RESULTS}Numerical Results}

The lattice simulations are carried out in the quenched approximation 
at $\beta = 2.230$ on a $12^3\times 24$ lattice using an 
improved Iwasaki gauge action. 
Hopping parameters  $\kappa_{ud}=0.1560$, $0.1580$, and $0.1600$
and $\kappa_s=0.1570$ are adopted for these simulations. 
The lattice spacing $a$ obtained using these parameters
corresponds to 0.8144 GeV$^{-1}$. 

Twenty different configurations separated by 2000 sweeps
are used to evaluate the correlation functions
$C_{\pi}$, $C_K$, and $C_{K\pi}$ at each $t$. 
We employ a complex $Z_2$ noise to represent the noise vectors in Eq.\ (\ref{eq:noise}).
The number of noise vectors in this equation is set to be four when the color and Dirac indices are fixed.

One could extend Eq.\ (\ref{eq:noise}) to include the color and Dirac 
indices. In this case, the  computational time  is reduced significantly, 
but obtained results suffer from large errors.
The choice here can be considered as a kind of dilution \cite{Foley05}.

\begin{figure}[htb]
\centerline{
\includegraphics[width=.7 \linewidth]{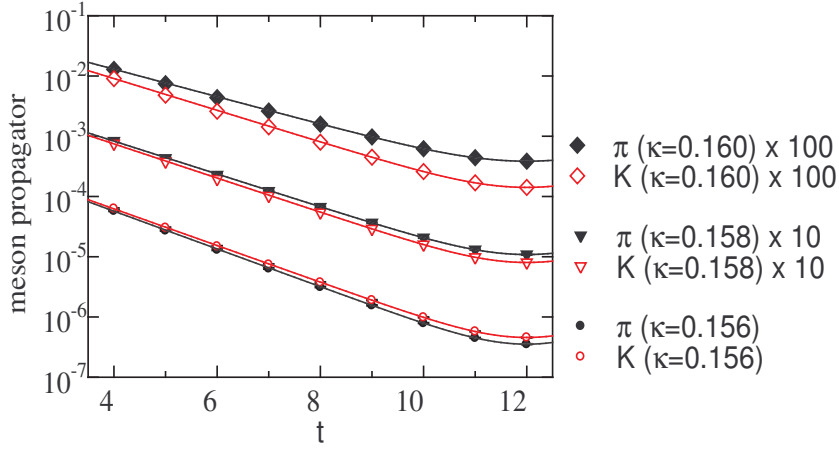}
}
\caption{
$C_{\pi}$ and $C_K$ for $\kappa_q = 0.156$, 0.158, and 0.160. 
\label{fig:propaKpi} 
}
\end{figure}

The obtained $C_{\pi}$ and $C_K$ are shown in Fig.\ \ref{fig:propaKpi}. Both 
these correlations are well reproduced by one-pole fitted functions
for $9 \le t \le 12$,
\be
C_{i}= Z_{i}\cosh(m_{i}(t-N_t/2)),
\ee 
with $i$ being $K$ or $\pi$. The obtained parameters 
are shown in Table \ref{table:meson-propa}.

\begin{table}[hbt]
\caption{Parameters for meson propagators} 
\label{table:meson-propa}
\begin{center}
\begin{tabular}{c|c|c}
\hline \hline 
 & $Z_{i}$ & $m_{i} a $\\ \hline
$C_{\pi}$ ($\kappa_q =0.156 $) & ( 0.35369  $\pm$ 0.0002283 ) $\times 10^{-6}$ & 0.722183 $\pm$ 0.0004114 \\ \hline
$C_K$ ($\kappa_q =0.156 $) & ( 0.457281 $\pm$ 0.0003214 ) $\times 10^{-6}$ & 0.701035 $\pm$ 0.0004315 \\ \hline
$C_{\pi}$ ($\kappa_q =0.158 $)&( 1.08796 $\pm$ 0.0008838 )  $\times 10^{-6}$ & 0.628463 $\pm$ 0.0004891 \\ \hline
$C_K$ ($\kappa_q =0.158 $)&
( 0.803039 $\pm$ 0.0005946 ) $\times 10^{-6}$ & 0.653207 $\pm$ 0.0004654 \\ \hline
$C_{\pi}$ ($\kappa_q =0.160 $) &
( 3.85386 $\pm$ 0.003642 ) $\times 10^{-6}$ & 0.526125 $\pm$ 0.0005567 \\ \hline
$C_K$ ($\kappa_q =0.160 $) &
( 1.41634 $\pm$ 0.001258 ) $\times 10^{-6}$ & 0.606525 $\pm$ 0.0005159 \\
\hline
\end{tabular}
\end{center}
\end{table}

\subsection{\label{sec:Diagrams}Diagrams $A, H$, and $X$}

\begin{figure}[htb]
\centerline{
\includegraphics[width=.7 \linewidth]{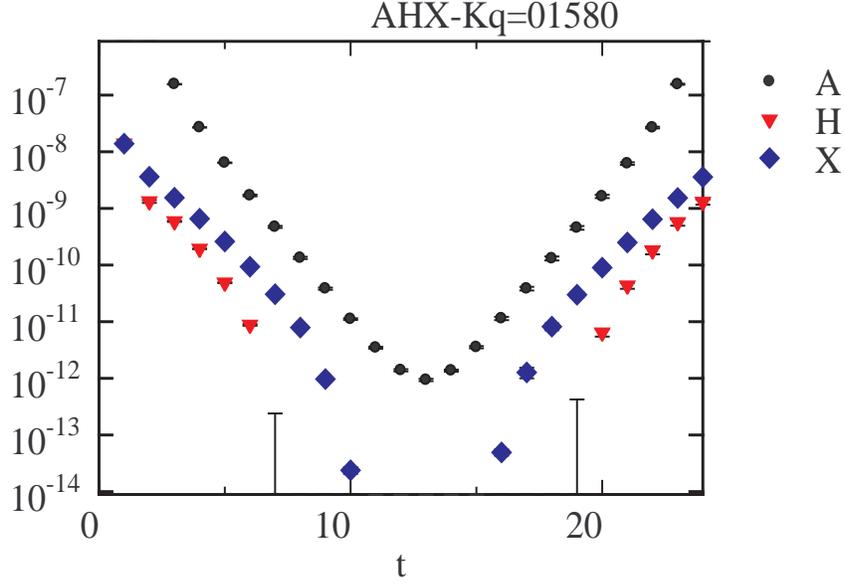}
}
\caption{
Numerical results for  $A$, $H$, and $X$ contributions in the $K\pi$
four-point function at $\kappa_q=0.158$.
\label{fig:AHXdata} 
}
\end{figure}

As shown in Sec.\ II, only three different diagrams contribute to the correlation
functions in the present system. Figure \ref{fig:AHXdata} shows 
the results obtained for  diagrams $A$, $H$, and
$X$ for $u$ and $d$ quarks of which hopping parameter,  $\kappa_q = 0.1580$. 
The $X$ and $H$ diagrams become negative at large $t$.
At higher values  of $\kappa_q=0.160$, the $H$ diagrams show
similar behaviors. 
Since the contributions of  $H$ and $X$ diagram
distinguish $I=1/2$  from $I=3/2$, precise measurement of these contributions 
in large $t$ regions is important.

\subsection{Lattice artifact}

\begin{figure}[htb]
\centerline{
\includegraphics[width=0.8 \linewidth]{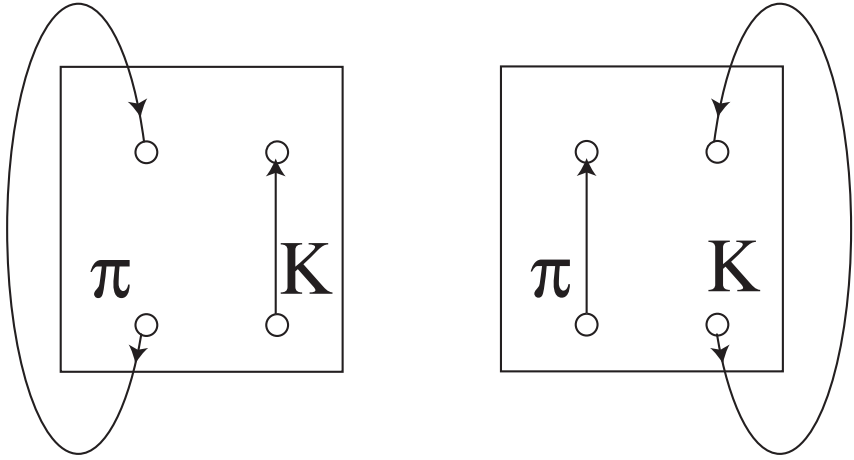}
}
\caption{ Diagrams that give rise to fake effects.
\label{fig:LatArtFact}
}
\end{figure}

\begin{figure}[bth]
\includegraphics[width=.5 \linewidth]{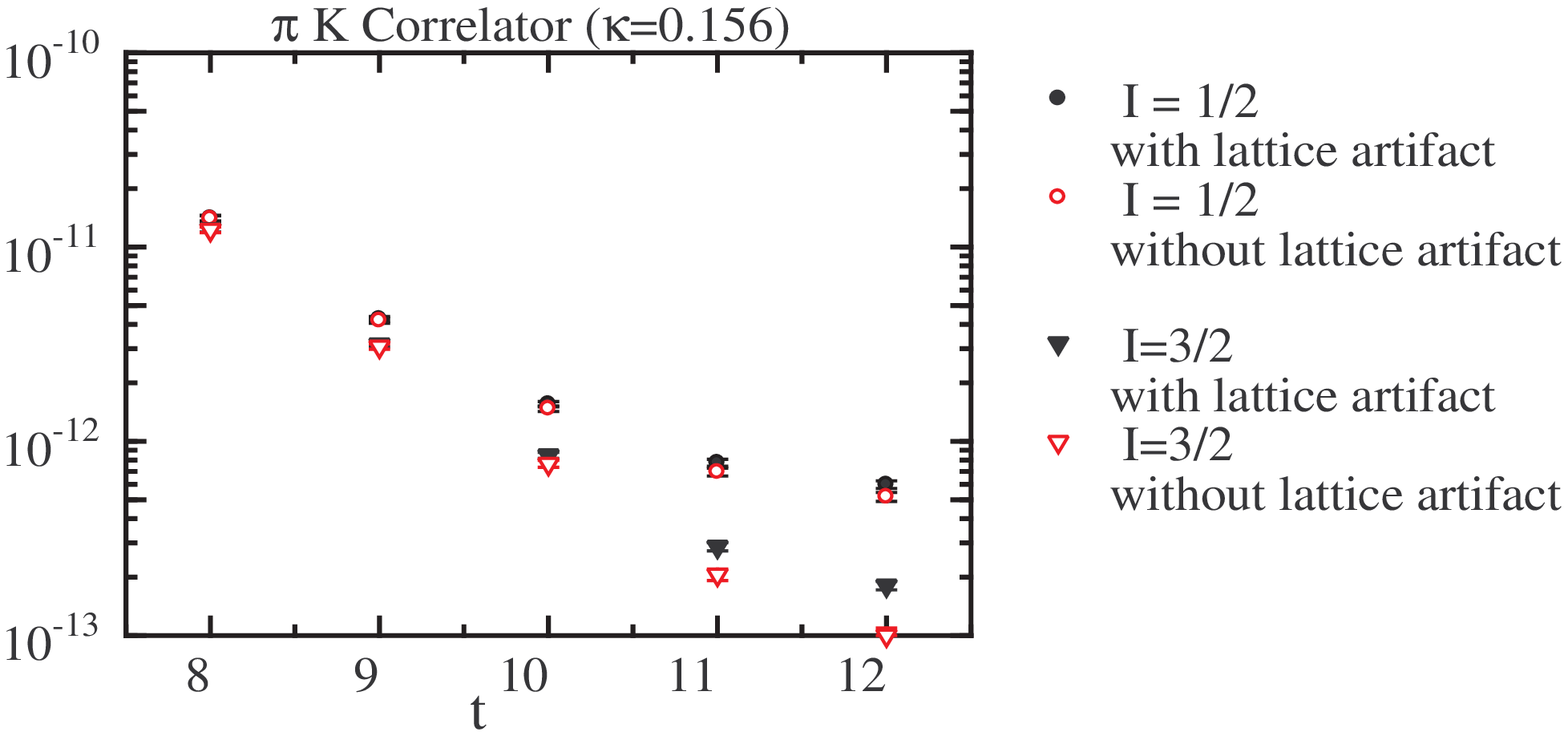}
\includegraphics[width=.5 \linewidth]{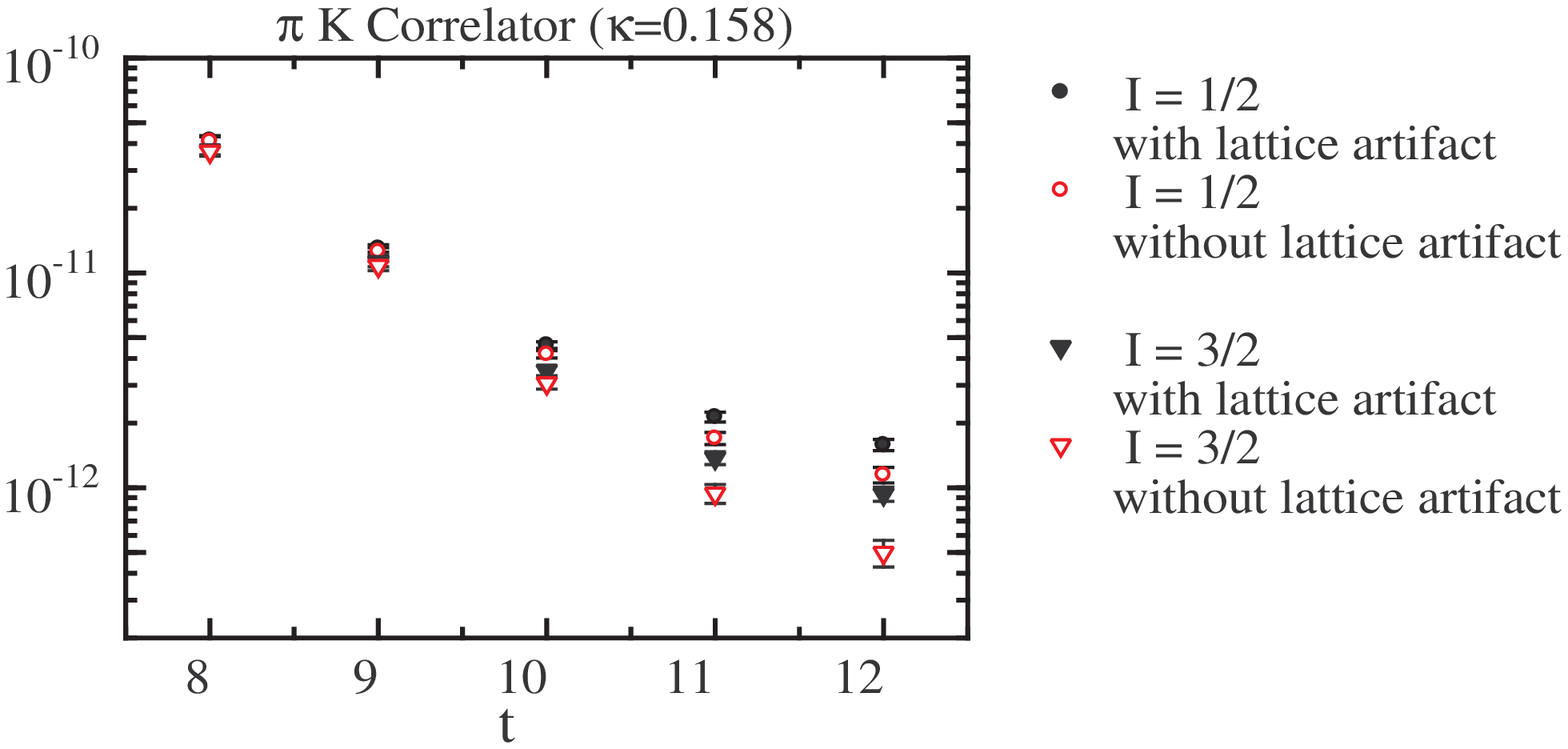}
\includegraphics[width=.5 \linewidth]{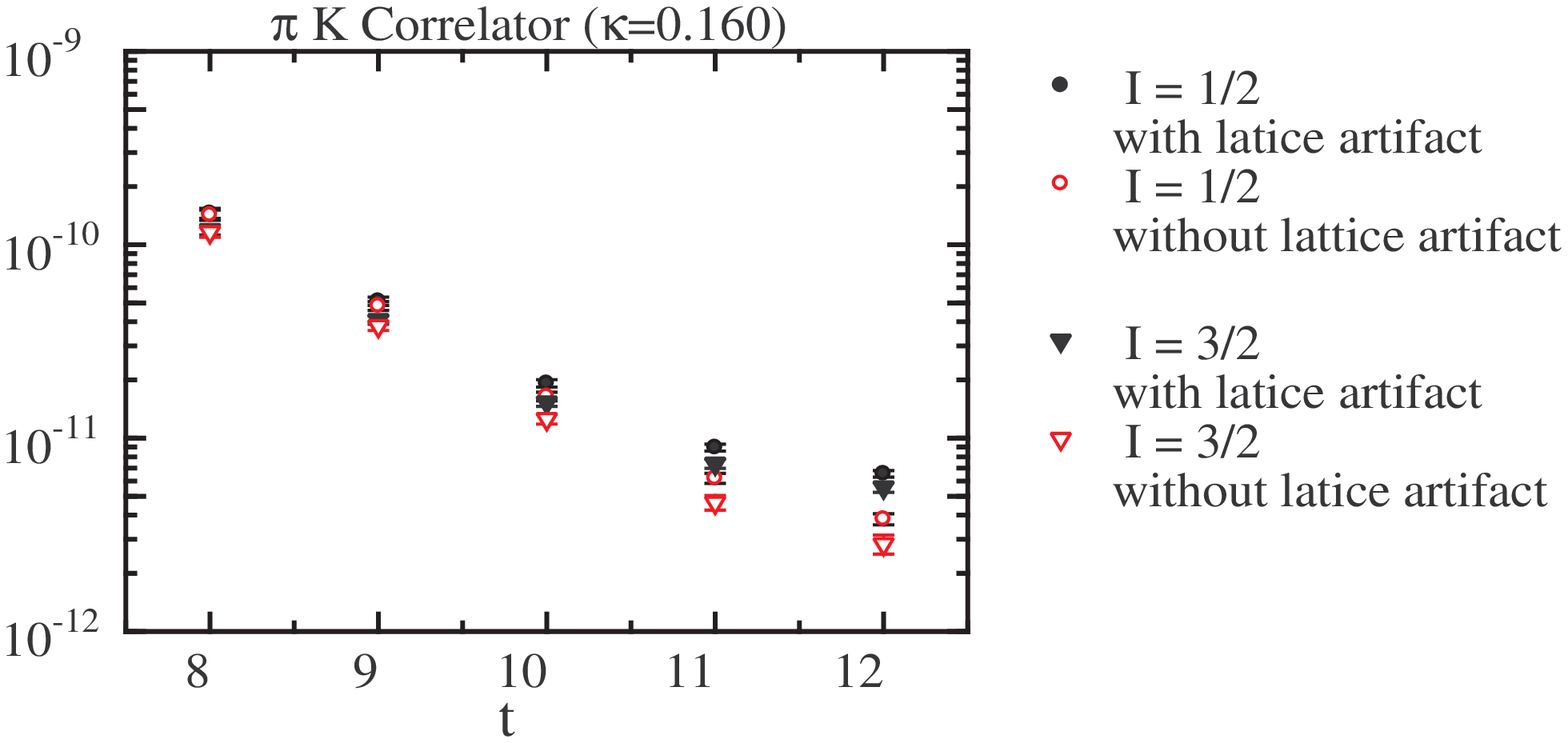}
\caption{
$K\pi$ four-point functions before and after subtracting
the lattice artifact, Eq.\ (\ref{Eq:pi-K}). 
The figures from top to bottom correspond to $K_q$ values
of 0.1560, 0.1580, and 0.1600, respectively. In each figure, for both $I=1/2$ 
and 3/2 channels, naive four-body correlations, $K\pi$, and those
obtained after correction for the  lattice artefact, Eq.\ (\ref{Eq:pi-K}), are shown.
\label{fig:LatArtFact2}
}
\end{figure}

Because the lattice extent is finite,  
our $K\pi$ amplitude contaminated with 
the artifact shown in Fig.\ \ref{fig:LatArtFact}.
These diagram contributions are large in meson correlator
calculations at finite temperature
in lattice QCD, and may mislead us as pointed out in Ref.\ \cite{Umeda07}. Such diagram are also
seen in the current lattice QCD simulations of the two-meson state.
This can be easily seen by evaluating the contribution
of the fake diagrams in Fig.\ \ref{fig:LatArtFact},
\bea
A_\pi  e^{-m_\pi (N_t - t)} \times A_K e^{-m_K t}
+
A_\pi  e^{-m_\pi  t} \times A_K e^{-m_K (N_t-t)}
\nonumber \\
= 2A_\pi A_K e^{-m_\pi N_t/2} e^{-m_K N_t/2} 
\cosh ((m_K-m_\pi)(t-N_t/2)) .
\label{Eq:pi-K}
\eea
When the mass difference between $\pi$ and $K$ is small, it acts as a
constant mode and distorts the four-point function at large $t$.
In Fig.\ \ref{fig:LatArtFact2}, we show the  above-mentioned contribution, 
Eq.\ (\ref{Eq:pi-K}),
together with the numerical data corresponding to  the four-point
functions.

Another method of avoiding the above mentioned fake diagram is 
to impose a Dirichlet boundary conditions.   
But here we use a simpler method than it ; we subtract 
the contribution of these diagrams numerically from
the obtained quantity.
$A_\pi$, $A_K$, $m_\pi$, and $m_K$ in Eq. (20) can be 
evaluated  from $\pi$ and $K$
propagator measurements.   In the present case, we have 
the data corresponding to  these two-point functions
are sufficiently  precise to  allow subtractions of these effects, and hence, Eq. 
(\ref{Eq:pi-K}) can be subtracted  from the $K \pi$ four-point data.
In the following fitting processes, the $K\pi$ four-point data is used 
after the subtraction.

\subsection{\label{sec:FITTING}Fitting analyses}


In general, propagators $C_{i}$ are composed of many excited states 
of the same quantum number.  
If $t$ is sufficiently large $(1 << t << N_t)$ and the
excited states have significantly 
higher masses than the ground-state mass,
the contribution of 
the lowest energy state is predominant in the case of 
this propagator, and the one-pole model,
\begin{equation}
C_{i}(0,t)=Z_{1}\cosh(-m_{1}(t-N_{t}/2) ), 
\label{one-pole}
\end{equation}
fits the numerical data well.
If there are contributions from higher states as well,
a two-pole model,
\begin{equation}
C_{i}(0,t)=Z_{1}\cosh(-m_{1}(t-N_{t}/2) ) + Z_{2}\cosh(-m_{2}(t-N_{t}/2) ),
\label{two-pole}
\end{equation}
would be more suitable than the one-pole model. 
Here, $m_{1}<m_{2}$, where $m_1$ 
is the lowest mass, and $m_2$ represents contributions from higher states.
If the contribution from the higher states is very small,
the fitting procedure for the two-pole model becomes unstable.

For $K$ and $\pi$ two-point functions, 
the one-pole ansatz, Eq.(\ref{one-pole}), 
works well, and hence,the masses of $K$ and $\pi$ are 
obtained with high accuracy (Fig.\ \ref{fig:propaKpi}).
The results of our simulation have already been  shown in 
 Table \ref{table:meson-propa}, and the values are consistent 
with those provided by CP-PACS. 

For calculating the propagators of the $K\pi$ system (meson four-point function), 
we adopt a two-pole model by 
taking into account higher excited states; 
however, the calculation method in this case is not simple.  
A naive application of Eq.\ (\ref{two-pole}) is not stable when 
the fitting region we use is changed. 
In order to obtain reliable results, we take the steps as follows:
\begin{enumerate}
\item
We apply Eq.\ (\ref{two-pole}) to $C_{K\pi}$ by changing the fitting
region.
\item
We choose a stable region, where the obtained $m_1$ shows
a plateau, and calculate the average of $m_{1}$ in this region.
\item
We also apply one-pole fitting and verify that $m_1$ obtained by
two-pole fitting is lower than that obtained by one-pole fitting.
\item
If the fitting procedure for the two-pole model is unstable, 
we adopt the results obtained with the one-pole model.
\end{enumerate}

\begin{figure}[htb]
\centerline{
\includegraphics[width=0.8 \linewidth]{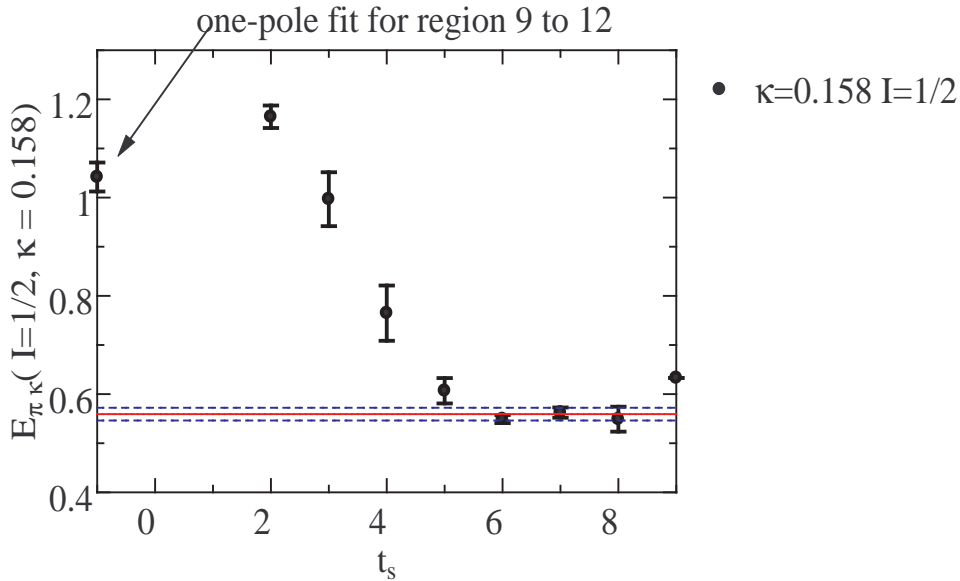}
}
\caption{Effective mass extracted by  one-pole and two-pole
fittings as a function of $t_S$. $I=1/2$ and $\kappa=0.158$.
The central region $(4 \le t_s \le 8)$ is used to determine
the ground-state energy. The value obtained and its error
are shown by horizontal solid and dashed lines, respectively.
The data point on the vertical axis shows the result obtained by the 
one-pole fit.
\label{fig:EffMass}
}
\end{figure}

In the present simulation, $N_t$ is 24 and 
the source field is set at $t = 0$. 
Because of its bosonic property, the propagator is symmetric at 
$t = N_t/2 = 12$.  
We apply the two-pole fit to  propagator $C_{K\pi}$ in the region
$t_{s} \le t \le N_t/2$.  
The larger $t_{s}$ corresponds to a propagator in the larger $t$ range, 
which is more reliable for picking up the ground state.  
However, in this case the number of available lattice points decreases.  
Figure \ref{fig:EffMass} shows $m_{1}$ as a function of $t_{s}$.  
Then, we obtain the statistical average of the 
plots in Fig.\ \ref{fig:EffMass} in order to obtain  our final results, 
which are shown as horizontal lines in the figure.

For $I= 3/2$ with $\kappa=$ 0.1560 and 0.1580, two-pole 
fitting yields results with large statistical errors. 
Therefore, in these cases, we adopt one-pole fitting in the region 
$9 \le t \le 12$.  
In the other cases, two-pole fitting gives better results than 
one-pole fitting, i.e., $m_1$ is smaller than the mass 
obtained with the one-pole model, 
and the statistical error is sufficiently small.
The obtained values of $E_{K\pi}$ and 
$\Delta E=E_{K\pi}-(m_{\pi}+m_{K})$ 
at different values of the hopping parameter $\kappa$ are 
summarized in Table \ref{tab:table1}.

\begin{table}
\caption{\label{tab:table1} Values obtained by fitting procedures
for four-point function $C_{K\pi}$ and two-point functions $C_K$ and $C_{\pi}$. 
The $S$-wave scattering length $a_{0}$ is also calculated using these values.
}
\begin{ruledtabular}
\begin{tabular}{c|c|ccc|c}
&$\kappa_{ud}$ & $E_{K\pi}$ & $\delta E_{K\pi}$ & $\Delta E$ &$a_0$ \\
\hline
       &0.1560 & 0.6542  & 0.01804  & -0.7690 & 0.4532 \\
I = 1/2&0.1580 & 0.9264  & 0.01882  & -0.3552 & 0.1598\\
       &0.1600 & 1.066   & 0.008285 & -0.0672 & 0.0222\\
\hline

       &0.1560 & 1.379 \footnote[1]{One-pole fit}&  0.02503 &  -0.0446 & 0.01850\\
I = 3/2&0.1580 & 1.258 $^a$             &  0.03848 &  -0.0240 & 0.00896\\
       &0.1600 & 1.087                         & 0.002129 &  -0.0453 & 0.01505\\
\end{tabular}
\end{ruledtabular}
\end{table}

\begin{figure}[htb]
\centerline{
\includegraphics[width=0.8 \linewidth]{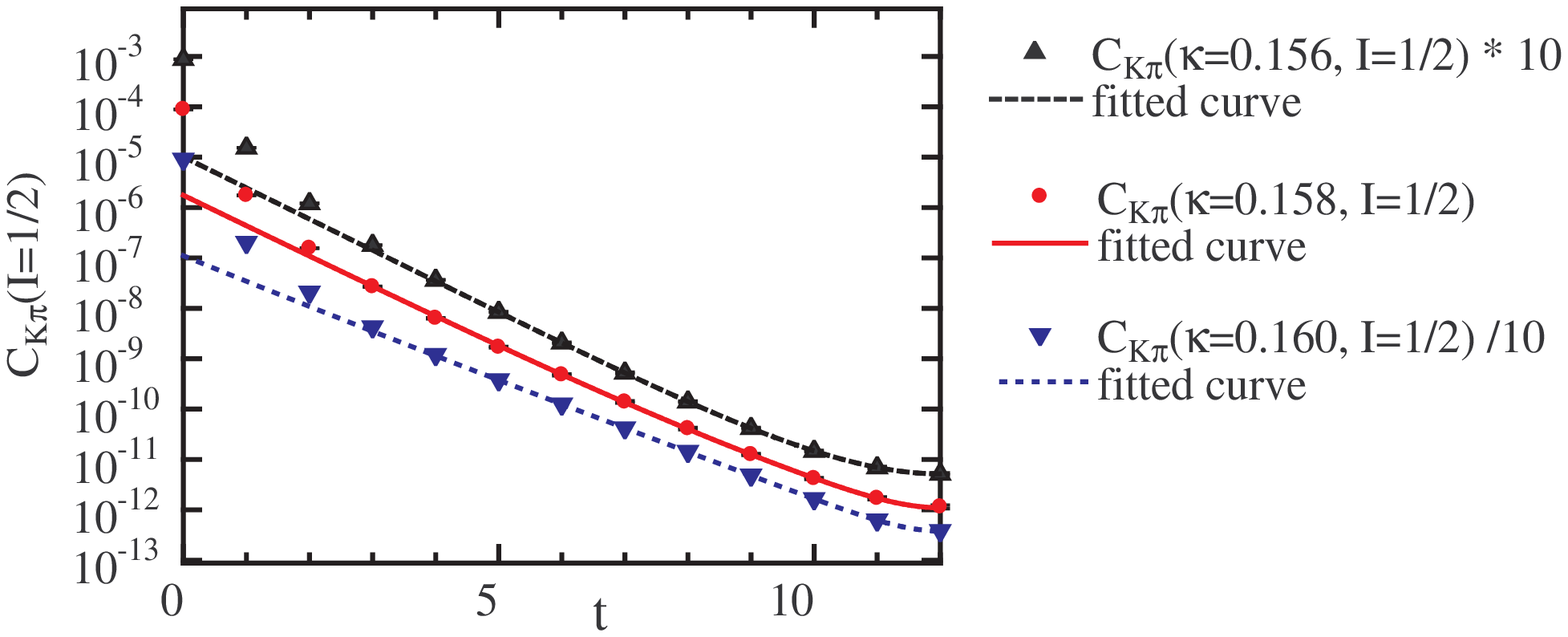}
}
\centerline{
\includegraphics[width=0.8 \linewidth]{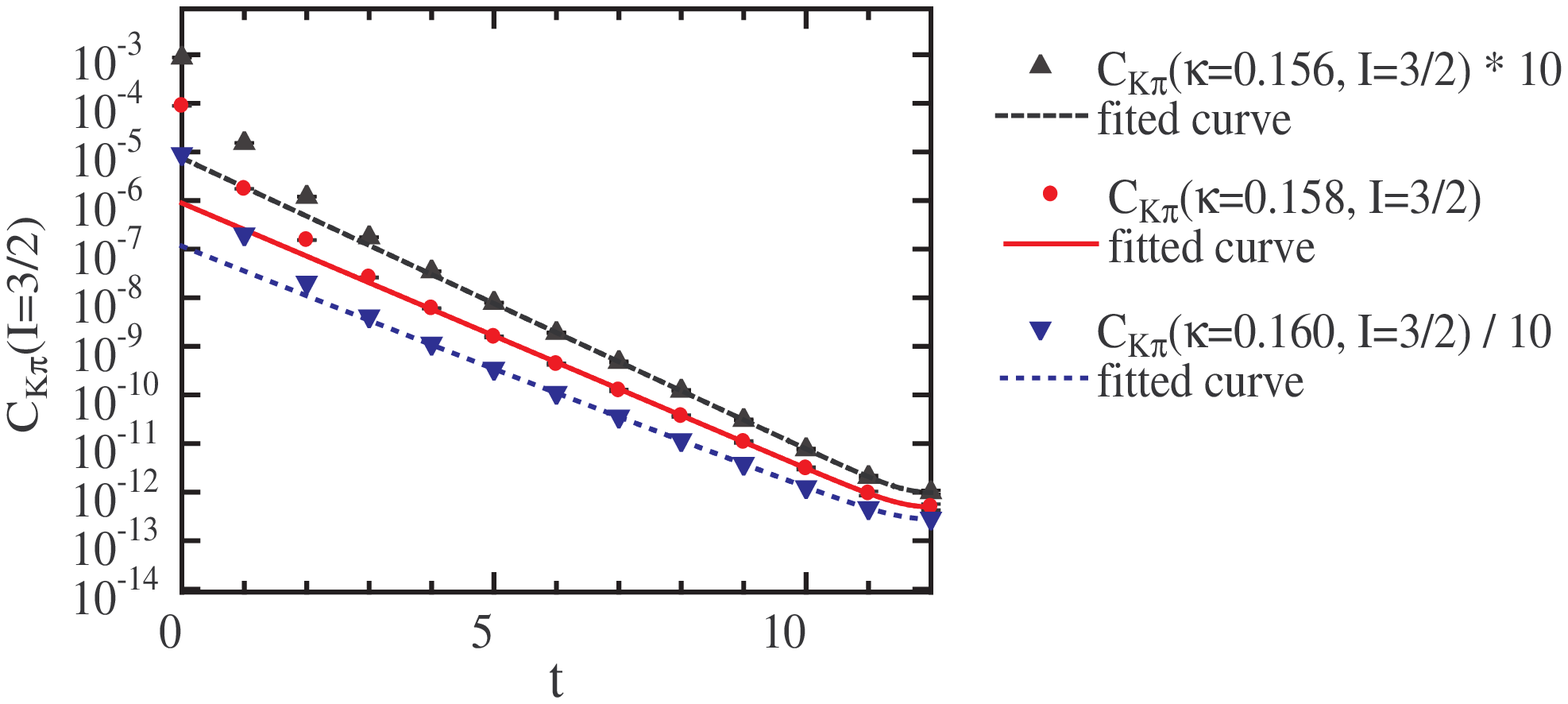}
}


\caption{Four-body correlator (propagators of the $K\pi$ system ) 
and fitting curves for different $\kappa$ values.   
Upper and lower figures correspond to the $I= 1/2$ and 
 $I=3/2$ channels, respectively.}
\label{fig:four-body}
\end{figure}


\subsection{\label{sec:CHIRAL}Chiral extrapolations and scattering length}

\begin{figure}[htb]
\centerline{
\includegraphics[width=0.8 \linewidth]{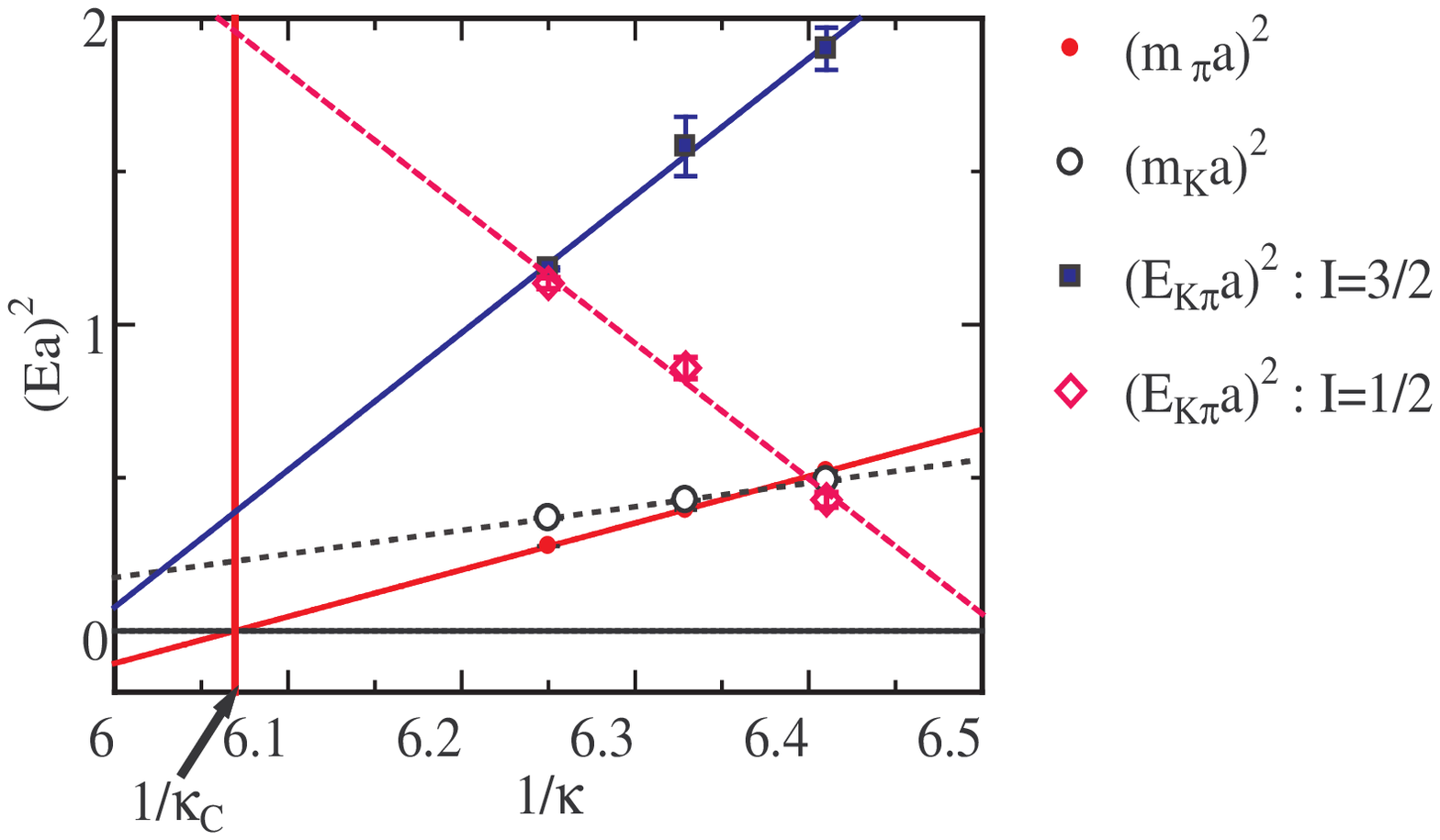}
}

\caption{ Chiral extrapolations of $m_\pi^2$, $m_K^2$, and
$E_{K\pi}^2$. The horizontal axis is $1/\kappa_q$, where
$\kappa_q$ is the hopping parameter for $u$ and $d$ quarks.
The vertical line represents 1/$\kappa_c$ = 6.069351464.
\label{fig:ChiralExtrapolation}
}
\end{figure}

In the present study, the pion masses $m_{\pi}$ are 
considerably larger than those determined experimentally, 
and hence,  we need to adopt an
extrapolation procedure.  Although the behavior of $E_{K\pi}^2$
near the chiral limit is not very clear, 
we plot $E_{K\pi}^2$ for $I=1/2$ and 3/2 as a function of $1/\kappa$ 
in Fig.\ \ref{fig:ChiralExtrapolation} 
together with $m_\pi^2$, and $m_K^2$; this is because $E_{K\pi}$ 
is expected to be dominated by
$m_\pi$ and $m_K$.

The least mean square procedure for a linear function, 
$f(x) = ax + b$, provides us the errors for both $a$ and $b$.  
The least square procedure assumes that the fitted line passes 
through the center of the weight of plotted points.  Therefore, 
errors of $a$ and $b$ are not independent but they are correlated; 
in the case that $a$ fluctuates positively $b$ must fluctuates negatively 
and vice versa. Hence, we may denote the line $y = a x +b$ as the middle 
line, $y = (a +\Delta a)x +(b-\Delta b)$ as the lower line and 
$y = (a-\Delta a)x+(b+\Delta b)$ as the upper line, with $\Delta a$ 
and $\Delta b$ being the error of $a$ and $b$, respectively. 
With these three lines, we evaluate the errors in 
the extrapolation  process.
Putting physical pion mass as $m_\pi =$ 139.57 MeV, the fitted line of 
$m_\pi ^2$ provides us hopping parameter at physical point as 
$1/\kappa_{phys}$=$6.077724 \pm 0.003071$. Here, the error,  $\Delta(1/\kappa_{phys})$ = 0.003071, is one half 
of the difference between the value of the upper line 
and the one of the lower line at the physical pion mass. 
Also in the case of the energy of the two-particle state, we denote three 
curves as the middle, the upper and the lower which are the square root 
of the corresponding lines of the linear fit to the square of the energy 
of the two-particle state, respectively. 
The central value at physical points is determined by the middle curve and $1/\kappa_{phys}$.  The error originates from the linear fitting, 
$\Delta_1$, is estimated from the difference between the upper curve and the lower curve at $1/\kappa_{phys}$.  There exists another kind of the error, $\Delta_2$, which originates from $1/\kappa_{phys}$ which is evaluated from the difference of the middle curve at $1/\kappa_{phys}+\Delta(1/\kappa_{phys})$ and at  $1/\kappa_{phys}-\Delta(1/\kappa_{phys})$.  As a final error value, we may take 
root mean square
of two independent errors, $\Delta_1$ and $\Delta_2$, as $\Delta E(\pi,K) = \sqrt{\Delta_1 ^2+\Delta_2 ^2}$.  We also 
estimate errors of $ E(\pi,K)$ in the other channel and $m_K$ in the same manner.


\begin{figure}[htb]
\centerline{
\includegraphics[width=0.8 \linewidth]{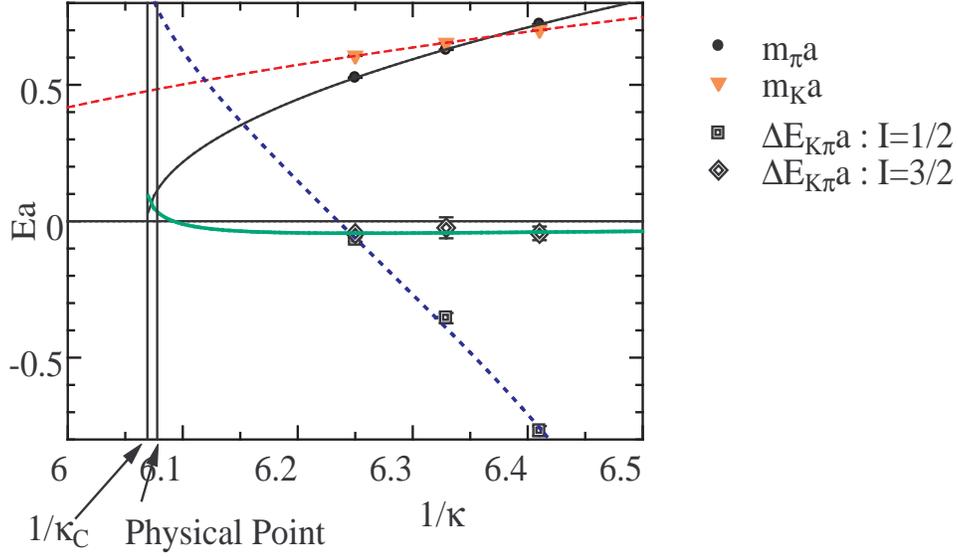}
}
\caption{$\Delta E=E_{K\pi}-(m_\pi+m_K)$ as a function of $1/\kappa_q$.
\label{fig:DeltaE}
}
\end{figure}

Substituting the obtained values into 
L\"{u}scher's formula, Eq. (\ref{luesher}), 
we can obtain scattering length $a_0$.  
Unfortunately, the results obtained in the present study 
are limited to a fixed volume; nevertheless,
by using the physical
size of the lattice unit, we can evaluate $a_{0}$.

\begin{figure}[htb]
\centerline{
\includegraphics[width=0.48 \linewidth]{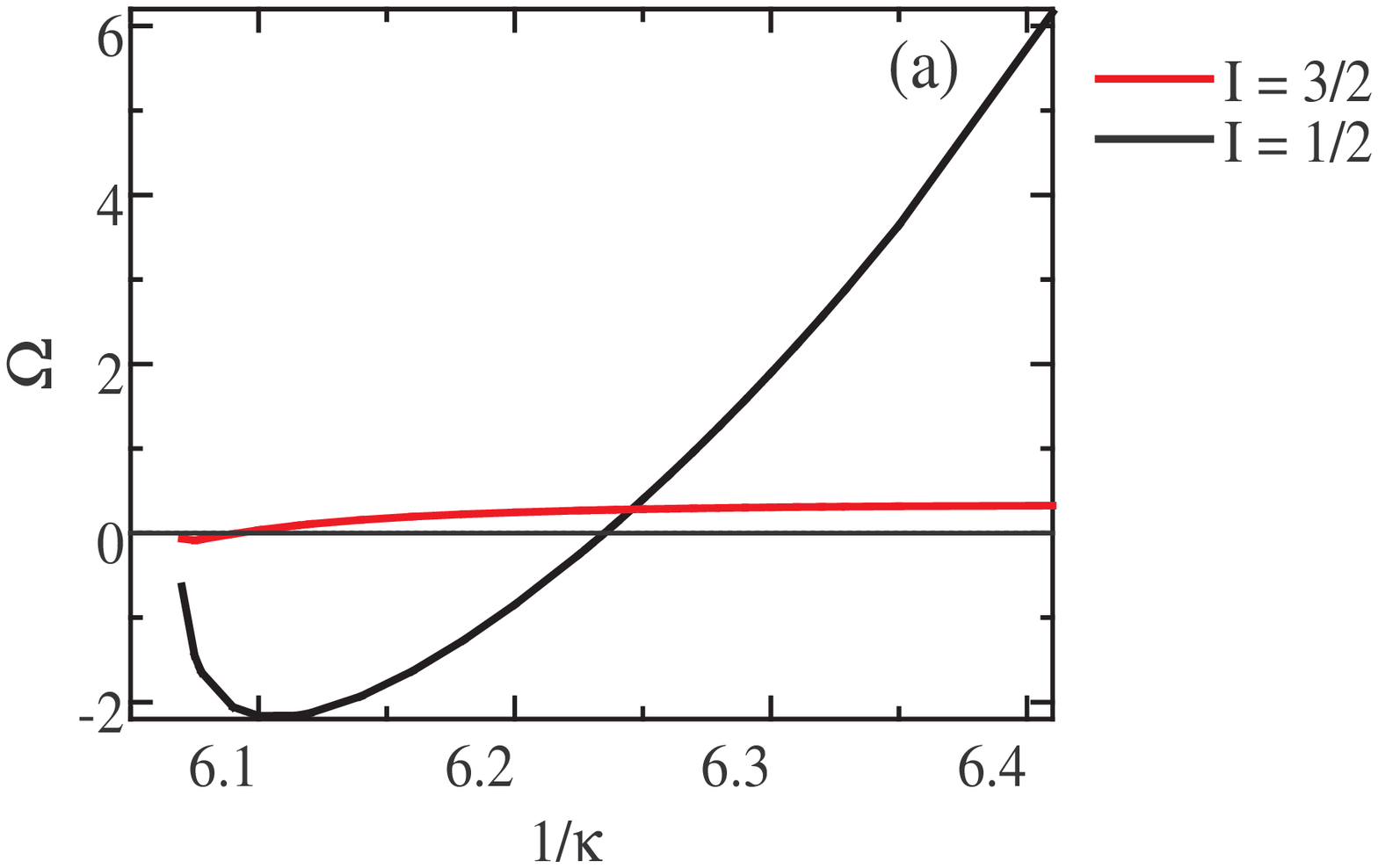}
\includegraphics[width=0.42 \linewidth]{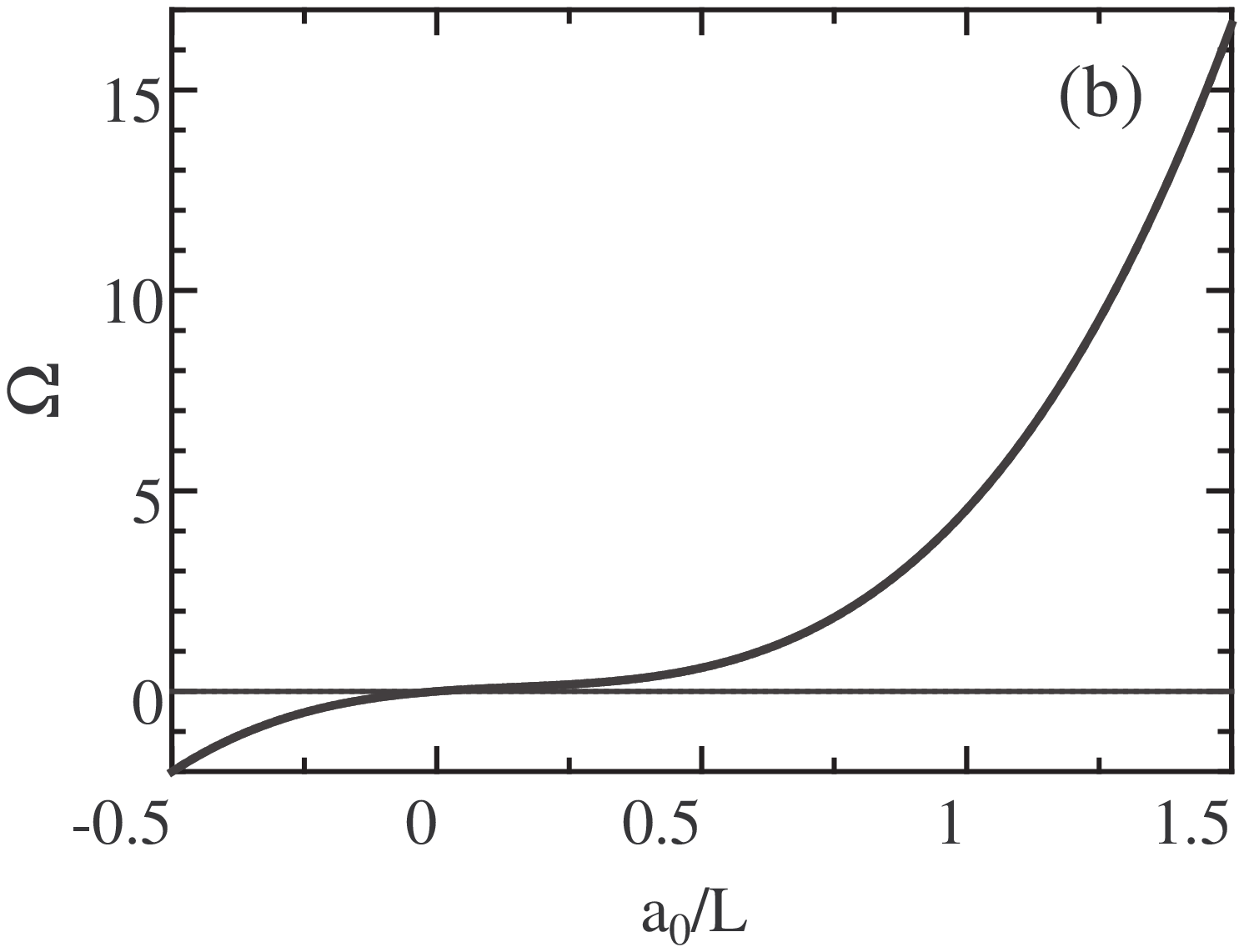}
}
\caption{$\Omega$ evaluated through $\Delta E$ obtained by fitted curve in Fig.\ \ref{fig:DeltaE} (left-hand side) and $\Omega$ as a function of $a_0/L$ (right-hand side).  
\label{fig:Omega}
}
\end{figure}

Let us rewrite Eq. (\ref{luesher}) as
\be
(E_{K \pi} - (m_K + m_{\pi}) )\times
\frac{ m_\pi m_K L^2}{2 \pi (m_K + m_{\pi})} = \Omega\left(\frac{a_0}{L}\right),  \\
\label{eq:luesher2a}
\ee
where
\be
\Omega\left(\frac{a_0}{L}\right) = \frac{  a_0}{ L }
           \bigl[ 1 + c_1 { a_0 \over L} + c_2 ( {a_0 \over L} )^2 \bigr] + O(L^{-6}).
\label{eq:luesher2b}
\ee
The left-hand side of Eq.\ (\ref{eq:luesher2a}) represents 
the shift in energy caused by the interaction between $\pi$ 
and $K$; this energy shift can be evaluated from the correlations.
The right-hand side, $\Omega\left(\frac{a_0}{L}\right)$, 
represents the effect of $a_{0}$ in the unit of box length. 
Figure \ref{fig:Omega} shows $\Omega$ as a function of $a_0 / L$.
$\Omega$ changes slowly at around ${a_{0} / L} \sim 0$, 
indicating that a small $a_0$ is sensitive to $\Delta E$ 
at around $\Delta E \sim 0$.
Further, $a_0$ easily changes from a positive small value to a negative small 
value and vice versa.

\begin{figure}[htb]
\centerline{
\includegraphics[width=0.8 \linewidth]{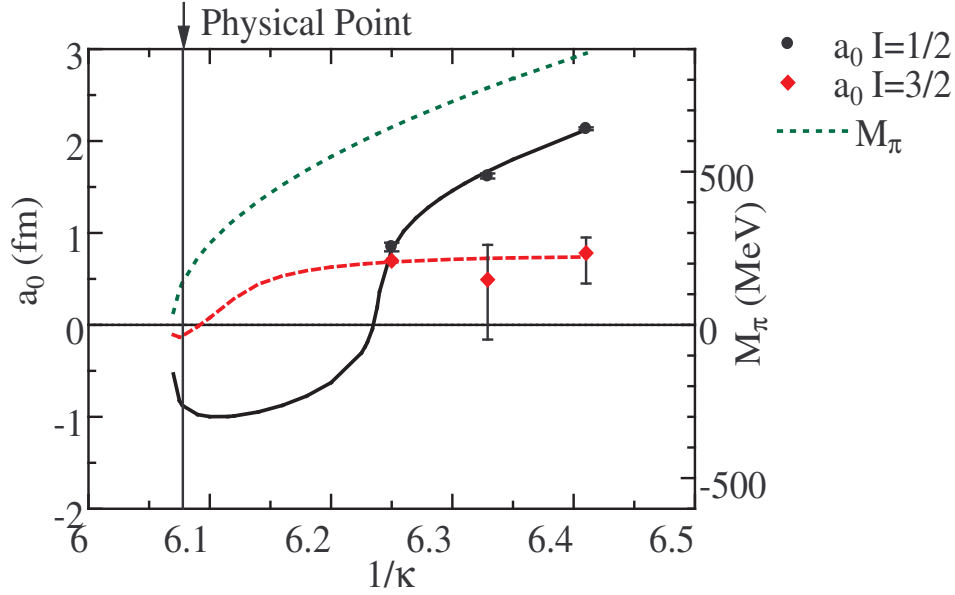}
}
\caption{S-wave scattering length $a_0$ as a function of $1/\kappa_q$.
Simulation data points at three values of $\kappa_q$ and
extrapolated values at $m_\pi=140$ MeV.
\label{fig:a0}
}
\end{figure}

The chiral extrapolation of $a_0$ is shown in Fig. \ref{fig:a0}. Our final results are summarized in Table \ref{tab:final}.
At all the calculation points, $a_0$ is positive, as shown in Table \ref{tab:table1}; however, the chiral extrapolation for $E_{K\pi}^2$
scaling leads to a change in the sign of $\Delta E$ and a subsequent
change in the sign of $a_0$.
At the physical point where $m_{\pi}=139.75$ MeV, $a_0(I=3/2)$ $m_\pi = -0.0837$ and $a_0(I=1/2)$ $m_\pi=-0.6248$.  In both cases, $\Delta E  > 0$, the phase shift $\delta <0$, and forces are repulsive.

\begin{table}
\caption{\label{tab:final} $a_{0}$ at the physical point where $m_\pi$=139.75 MeV.}
\begin{center}
\begin{tabular}{ccccc}
\hline
\hline
 & & {$a_0$}(MeV) & &$a_0$ $m_{\pi}$ \\
\hline
$I=1/2$   && -0.8794 $^{+ 0.016}_{ - 0.017}$  
 && -0.6248 $^{+ 0.0115}_{- 0.0118}$  \\     
$I = 3/2$ &&  -0.1178 $^{+ 0.0712} _{- 0.0901}$ 
&&  -0.0837 $^{+ 0.0506} _{ - 0.0640}$ 
 \\
\hline
\end{tabular}
\end{center}
\end{table}

\section{\label{sec:CONCLUSIONS}Concluding Remarks}

The scattering length, $a_0$ of the $I=3/2$ channel in $K\pi$ scattering 
has been  studied 
by theoretical and experimental approaches so far\cite{Joh73,Mat74,Kar80,Ber91a,Ber91b,Bue03,Lla04}.
Previous experiments have reported that $a_{3/2}m_{\pi}$ has a
small negative value, i.e.,  
$-0.13\sim -0.05$\cite{Joh73,Mat74,Kar80}. 
Small negative value was also claimed  by a theoretical model 
based on the chiral perturbation 
theory as $a_0 = -0.129 \sim -0.05$\cite{Ber91a,Ber91b,Bue03}. 
The first lattice calculation of $K\pi$ scattering 
in the $I=3/2$ channel was performed by
Miao {\it et al.}\cite{Mia04}, and the value of $a_0$ was found
to be $-0.048$.    

On the other hand, no direct simulations have 
been carried out on 
$K\pi$ scattering in the $I=1/2$ channel
for estimations of $a_0$.  
The NPLQCD group carried out lattice simulations for $K\pi$ scattering
in the $I=3/2$ channel and evaluated the $a_0$ values for 
both $I=3/2$ and $1/2$; the results of $I=1/2$ were obtained on the basis 
of chiral perturbations and not directly by simulations 
\cite{Bea06}. 
They obtained a small negative value of $-0.0574$ for the
$I=3/2$ channel and found that 
$m_{\pi} a_{1/2}=+0.1725^{+0.0029}_{-0.0157}$ for the $I=1/2$ channel.
The pion mass $m_{\pi}$ used by them was lower than that used
in our present study. 
We could carry out a direct comparison of our results
with those of NPLQCD group if we carry out our simulations at a
low $m_\pi$.
Flynn and Nieves\cite{Fly07} used the scalar form factors in semi-leptonic 
pseudo-scalar-to-pseudo-scalar decays to extract information 
on $K\pi$ scattering
in the $I=1/2$ channel and obtained $m_{\pi} a_{1/2}$ = +0.179(17)(14).

In this paper, we presented a lattice QCD simulation of the 
$K\pi$ scattering length
and formulas that make this calculation simple and feasible.
We have observed the followings:
\begin{itemize}
\item
$I=1/2$ and $3/2$ amplitudes can be expressed using quark diagrams, $A$, $H$ and $X$.
\item
use of a reasonable noise vector works well.
\item
a simple method for eliminating fake diagrams caused by the finite size of lattice is useful.
\item
the  accurate measurements in large $t$ regions is important.
\item
the behavior of $\Delta E$ in the chiral extrapolation needs careful analysis.
\end{itemize}

Thus, it is now possible to study $K\pi$
scattering reactions on the basis of lattice QCD simulations.
However,
our present study does not provide adequate information on
scalar mesons with the strangeness, i.e., $\kappa$,
because our study is restricted to the zero-momentum case.
Therefore, this paper does not include results on the phase shift $\delta(p)$,
which is indicative of a  $\kappa$ pole in the $K\pi$ channel. 
The study of $K\pi$ interactions is the first  step 
in the study of hadron interactions including $s$-quarks, 
and we are now
entering into the new era to study hyperon interaction
from QCD.  

We show exchanged mesons for meson-meson and nucleon-hyperon interactions in 
Table \ref{tab:table3} and \ref{tab:table4}.
\footnote{The baryon table was constructed from
Eq.(17.8) of Ref.\cite{Swart63}. 
The absence of $\pi\Lambda\Lambda$ coupling is a direct
consequence of  iso-spin $I=0$ for $\Lambda$ particle.
}
$NN$ interactions include $\pi$ exchange and require lattices
of very large size for estimation of the its large scattering length.
On the contrary,
$N \Lambda$ interactions can be 
studied using a lattice of reasonable size.
Hence, 
$N\Lambda$ interactions 
are suitable for lattice studies because
they have interaction ranges that can be fit to the L\"uscher's formula;
these interactions will be extensively studied in future experiments
in J-PARC and other laboratories.

\hfil\break

\begin{acknowledgments}
We thank Dr.\ Chiho Nonaka for her sincere help,
and Prof. Takashi  Nakano for explaining us experimental situation of
$K\pi$ interaction.
Discussion with Prof.\ Naomichi Suzuki was fruitful.
This work was supported in part by a Grant-in-Aid for Scientific Research (C)
from JSPS (No. 17540272, 18540294, 17340080, 18540294, 20340055), 
and the Large Scale Simulation Program No.\ 06-05 (FY2006) 
of the High Energy Accelerator Research Organization (KEK).
SX-8 at RCNP, Osaka University, and SR1100 at Hiroshima University.
A part of analyses was done by computers in Matsumoto University.
\end{acknowledgments}

\begin{table}
\caption{\label{tab:table3} 
Exchanged pseudo-scalar mesons for baryon-baryon scatterings. 
}
\begin{tabular}{|l|c|}
\hline
\hline
scattering system & exchanged mesons \\
\hline
 $N - N$ & $\pi$, $\eta$, $\eta'$  \\
 $ \Lambda - N $ & $K$, $\eta$, $\eta'$  \\
 $\Sigma - N$ & $\pi$, $K$ \\
 $ \Xi - N $ & $\pi$, $\eta$, $\eta'$ \\
\hline
\end{tabular}
\end{table}

\begin{table}
\caption{\label{tab:table4} 
Main contributions in $t/u$- and $s$-channels  for pseudo-scalar
meson-meson scattering.
}
\begin{tabular}{|l|c|c|}
\hline
\hline
scattering system & $t/u$-channel & $s$-channel \\
\hline
 $\pi - \pi$ & $\rho$, $\sigma$  & $\rho$, $\sigma$ \\
 $K - \pi $ & $\rho$, $\sigma$  & $K^{*}$, $\kappa$ \\
\hline
\end{tabular}
\end{table}

\newpage
\appendix
\section{
\label{sec:AppendixA}
$I=1/2$  $K\pi$ scattering amplitudes in terms of quark propagators
}
\bea
&&\la O_{K^{0}}(\xop)O_{\pi^{+}}(\xtp) 
O_{K^{0}}^{\dagger}(\xo)O_{\pi^{+}}^{\dagger}(\xt)\ra \nonumber\\
&=& {\rm Tr}\left( G^{(d)}(x_{2},x_{2}') \gamma_{5} 
    G^{(u)}(x_{2}',x_{2}) \gamma_{5} \right) 
 {\rm Tr}\left( G^{(d)}(x_{1}',x_{1}) \gamma_{5} G^{(s)}(x_{1},x_{1}') \gamma_{5} \right) \label{term_01}
\\
&-& 
{\rm Tr} \left( G^{(d)}(x_{2},x_{1}) \gamma_{5} G^{(s)}(x_{1},x_{1}') \gamma_{5}
G^{(d)}(x_{1}',x_{2}') \gamma_{5} G^{(u)}(x_{2}',x_{2}) \gamma_{5} \right) 
\label{term_02}
\eea

\bea
&&\la O_{K^{0}}(\xop)O_{\pi^{+}}(\xtp) 
O_{K^{+}}^{\dagger}(\xo)O_{\pi^{0}}^{\dagger}(\xt)\ra \nonumber\\
&=&- \frac{1}{\sqrt{2}} 
{\rm Tr}\left( G^{(u)}(x_{2},x_{2}) \gamma_{5} \right)
{\rm Tr}\left( G^{(u)}(x_{2}',x_{1}) \gamma_{5} 
 G^{(s)}(x_{1},x_{1}') \gamma_{5} G^{(d)}(x_{1}',x_{2}') \gamma_{5} \right) \label{term_03}
\\
&& +{ 1 \over\sqrt{2} }
{\rm Tr}\left( G^{(u)}(x_{2}',x_{2}) \gamma_{5} G^{(u)}(x_{2},x_{1}) \gamma_{5} 
 G^{(s)}(x_{1},x_{1}') \gamma_{5} G^{(d)}(x_{1}',x_{2}') \gamma_{5} \right) \label{term_04}
\\
&&+ \frac{1}{\sqrt{2}} 
{\rm Tr}\left( G^{(d)}(x_{2},x_{2}) \gamma_{5} \right)
{\rm Tr}\left( G^{(s)}(x_{1},x_{1}') \gamma_{5} 
 G^{(d)}(x_{1}',x_{2}') \gamma_{5} G^{(u)}(x_{2}',x_{1}) \gamma_{5} \right) \label{term_05} \\
&& - \frac{1}{\sqrt{2}} 
{\rm Tr}\left( G^{(d)}(x_{1}',x_{2}) \gamma_{5} G^{(d)}(x_{2},x_{2}') \gamma_{5} 
 G^{(u)}(x_{2}',x_{1}) \gamma_{5} G^{(s)}(x_{1},x_{1}') \gamma_{5} \right) \label{term_06}
\eea

\bea
&&\la O_{K^{+}}(\xop)O_{\pi^{0}}(\xtp) 
O_{K^{0}}^{\dagger}(\xo)O_{\pi^{+}}^{\dagger}(\xt)\ra \nonumber\\
&=&- \frac{1}{\sqrt{2}} 
{\rm Tr}\left( G^{(u)}(x_{2}',x_{2}') \gamma_{5} \right)
{\rm Tr}\left( G^{(u)}(x_{1}',x_{2}) \gamma_{5} 
 G^{(d)}(x_{2},x_{1}) \gamma_{5} G^{(s)}(x_{1},x_{1}') \gamma_{5} \right) \label{term_07}
\\
&& +\frac{1}{\sqrt{2}} 
{\rm Tr}\left( G^{(u)}(x_{2}',x_{2}) \gamma_{5} G^{(d)}(x_{2},x_{1}) \gamma_{5} 
 G^{(s)}(x_{1},x_{1}') \gamma_{5} G^{(u)}(x_{1}',x_{2}') \gamma_{5} \right) \label{term_08}
\\
&&+ \frac{1}{\sqrt{2}} 
{\rm Tr}\left( G^{(d)}(x_{2}',x_{2}') \gamma_{5} \right)
{\rm Tr}\left( G^{(d)}(x_{2},x_{1}) \gamma_{5} 
 G^{(s)}(x_{1},x_{1}') \gamma_{5} G^{(u)}(x_{1}',x_{2}) \gamma_{5} \right) \label{term_09} \\
&& - \frac{1}{\sqrt{2}} 
{\rm Tr}\left( G^{(d)}(x_{2}',x_{1}) \gamma_{5} G^{(s)}(x_{1},x_{1'}) \gamma_{5} 
 G^{(u)}(x_{1}',x_{2}) \gamma_{5} G^{(d)}(x_{2},x_{2}') \gamma_{5} \right) \label{term_10}
\eea

\bea
&&\la O_{K^{+}}(\xop)O_{\pi^{0}}(\xtp) 
O_{K^{+}}^{\dagger}(\xo)O_{\pi^{0}}^{\dagger}(\xt)\ra \nonumber \\
&=&- \frac{1}{2} 
{\rm Tr}\left( G^{(u)}(x_{2}',x_{2}') \gamma_{5} \right)
{\rm Tr}\left( G^{(u)}(x_{2},x_{2}) \gamma_{5} \right)
{\rm Tr}\left( G^{(d)}(x_{1}',x_{1}) \gamma_{5} G^{(s)}(x_{1},x_{1}') \gamma_{5} 
\right) \label{term_11} \\
&&+ \frac{1}{2} 
{\rm Tr}\left(G^{(u)}(x_{1}',x_{1}) \gamma_{5} G^{(s)}(x_{1},x_{1}') \gamma_{5} \right)
{\rm Tr}\left( G^{(u)}(x_{2}',x_{2}) \gamma_{5} G^{(u)}(x_{2},x_{2}') \gamma_{5} \right) \label{term_12}
\\
&&- \frac{1}{2} 
{\rm Tr}\left( G^{(u)}(x_{1}',x_{2}') \gamma_{5} 
G^{(u)}(x_{2}',x_{2}) \gamma_{5} 
 G^{(u)}(x_{2},x_{1}) \gamma_{5} G^{(s)}(x_{1},x_{1}') \gamma_{5} \right) \label{term_13} \\
&& + \frac{1}{2} 
{\rm Tr}\left( G^{(u)}(x_{1}',x_{2}') \gamma_{5} 
G^{(u)}(x_{2}',x_{1}) \gamma_{5} 
 G^{(s)}(x_{1},x_{1}') \gamma_{5} \right)
{\rm Tr}\left( G^{(u)}(x_{2},x_{2}) \gamma_{5} \right) \label{term_14} \\
&& - \frac{1}{2} 
{\rm Tr}\left( G^{(u)}(x_{1}',x_{2}) \gamma_{5} G^{(u)}(x_{2},x_{2}') \gamma_{5} 
 G^{(u)}(x_{2}',x_{1}) \gamma_{5} G^{(s)}(x_{1},x_{1}') \gamma_{5} \right) \label{term_15} \\
&& + \frac{1}{2} 
{\rm Tr}\left( G^{(u)}(x_{1}',x_{2}) \gamma_{5} G^{(u)}(x_{2},x_{1}) \gamma_{5} 
 G^{(s)}(x_{1},x_{1}') \gamma_{5} \right)
{\rm Tr}\left( G^{(u)}(x_{2}',x_{2}') \gamma_{5} \right) \label{term_16} \\
&+& \frac{1}{2} 
{\rm Tr}\left( G^{(u)}(x_{1}',x_{1}) \gamma_{5} 
G^{(s)}(x_{1},x_{1}') \gamma_{5} \right)
{\rm Tr}\left( G^{(u)}(x_{2}',x_{2}') \gamma_{5}\right)
{\rm Tr}\left( G^{(d)}(x_{2},x_{2}) \gamma_{5} 
\right) \label{term_17} \\
&&- \frac{1}{2} 
{\rm Tr}\left(G^{(u)}(x_{1}',x_{2}') \gamma_{5} G^{(u)}(x_{2}',x_{1}) \gamma_{5} G^{(s)}(x_{1},x_{1}') \gamma_{5} \right)
{\rm Tr}\left(G^{(d)}(x_{2},x_{2}) \gamma_{5} \right) \label{term_18}
\\
&&+ \frac{1}{2} 
{\rm Tr}\left( G^{(u)}(x_{1}',x_{1}) \gamma_{5} 
G^{(s)}(x_{1},x_{1}') \gamma_{5} \right)
{\rm Tr}\left( G^{(u)}(x_{2},x_{2}) \gamma_{5} \right)
{\rm Tr}\left( G^{(d)}(x_{2}',x_{2}') \gamma_{5} \right) \label{term_19} \\
&& - \frac{1}{2}
{\rm Tr}\left( G^{(u)}(x_{1}',x_{2}) \gamma_{5} 
G^{(u)}(x_{2},x_{1}) \gamma_{5} 
 G^{(s)}(x_{1},x_{1}') \gamma_{5} \right)
{\rm Tr}\left( G^{(d)}(x_{2}',x_{2}') \gamma_{5} \right) \label{term_20} \\
&& - \frac{1}{2} 
{\rm Tr}\left( G^{(d)}(x_{2}',x_{2}') \gamma_{5} \right)
{\rm Tr}\left( G^{(d)}(x_{2},x_{2}) \gamma_{5} \right)
{\rm Tr}\left( G^{(s)}(x_{1},x_{1}') \gamma_{5} G^{(u)}(x_{1}',x_{1}) \gamma_{5} \right) \label{term_21} \\
&& + \frac{1}{2} 
{\rm Tr}\left( G^{(d)}(x_{2}',x_{2}) \gamma_{5} G^{(d)}(x_{2},x_{2}') \gamma_{5}\right) 
{\rm Tr}\left( G^{(s)}(x_{1},x_{1}') \gamma_{5} 
 G^{(u)}(x_{1}',x_{1}) \gamma_{5} \right) \label{term_22} 
\eea

\section{
\label{sec:AppendixB}
Explicit forms of the diagram $A$, $H$ and $X$}


In this section, we employ two independent random noises,
$\xi$ and $\eta$. They are a function of the spatial coordinate,
i.e., they live on a each time slice. 
\bigskip
\bea
\Sj
\xi_{j}(\vec{x})^{\dagger} \xi_{j}(\vec{y})
&=& \delta_{\vec{x},\vec{y}}
\\
\Sj
\eta_{j}(\vec{x})^{\dagger} \eta_{j}(\vec{y})
&=& \delta_{\vec{x},\vec{y}}
\eea

\subsection{Diagram A}

\bea
&A&(\vpn,-\vpn,\vpm,-\vpm) 
\\
&=& 
\Svxop \Svxtp \Svxo \Svxt
\FTpno \FTpnt \FTpmo \FTpmt
\\
&\times&
A((\xxop),(\xxtp),(\xxo),(\xxt))
\\
&=& \Svxop \Svxo \FTpno \FTpmo 
\\
&\times& \Tr\left(\gf \Gs(\xxo;\xxop) \gf G(\xxop;\xxo) \right)
\\
&\times& \Svxtp \Svxt \FTpnt \FTpmt
\\
&\times& \Tr\left(\gf G(\xxt;\xxtp) \gf G(\xxtp;\xxt) \right)
\\
&=& \Sjo
\Svyo \Svxop \Svxo \FTpno \FTpmo 
\\
&\times& \Tr\left(\xi_{j_1}^{\dagger}(\vyo) \gf \Gs(\yyo;\xxop) \gf 
    G(\xxop;\xxo) \xi_{j_1}(\vxo) \right)
\\
&\times& \Sjt \Svyt \Svxtp \Svxt \FTpnt \FTpmt
\\
&\times& \Tr\left(\eta_{j_2}^{\dagger}(\vyt) \gf G(\yyt;\xxtp) 
\gf G(\xxtp;\xxt) \eta_{j_2}(\vxt) \right)
\eea

We can write the trace terms as
\be
 \Tr\left(\gf \Gs(\xxo;\xxop) \gf G(\xxop;\xxo) \right)
= \sum_{a,\alpha} \la a,\alpha| 
 \gf \Gs(\xxo;\xxop) \gf G(\xxop;\xxo) |a, \alpha\ra 
\ee
\be
 \Tr\left(\gf G(\xxt;\xxtp) \gf G(\xxtp;\xxt) \right)
= \sum_{a,\alpha} \la a,\alpha| 
 \gf G(\xxt;\xxtp) \gf G(\xxtp;\xxt) |a, \alpha\ra 
\ee
where $a$ and $\alpha$ stand for the color and Dirac
indices, respectively.


Then
\bea
A(\vpn,-\vpn,\vpm,-\vpm;t,t_S) 
= 
\Sjo \sum_{a,\alpha}
\Svxop
\vec{Z}_1(\vxop,t;t_S)^{\dagger} \FTpno 
\vec{Y}_1(\vxop,t;t_S)
\nonumber \\
\times
\Sjt \sum_{a,\alpha}
\Svxtp
\vec{Z}_2(\vxtp,t;t_S) ^{\dagger} \FTpnt 
\vec{Y}_2(\vxtp,t;t_S)
\eea
where
\bea
\vec{Y}_1 
&\equiv& 
G \FTpmo \xi_{j_1}|a, \alpha\ra 
\quad\mbox{or}
\\
\vec{Y}_1(\vxop,t;t_S) 
&=&
\Svxo  G(\xxop;\xxo) \FTpmo 
\xi_{j_1}(\vxo) |a, \alpha\ra 
\eea
\bea
\vec{Z}_1 &\equiv& 
\gf \Gs^{\dagger}\gf \xi_{j_1} |a, \alpha\ra 
\quad\mbox{or}
\\
\vec{Z}_1(\vxop,t;t_S) &=& 
\Svyo   \gf \Gs^{\dagger}(\xxop;\yyo) \gf 
\xi_{j_1}(\vyo) |a, \alpha\ra 
\eea
\bea
\vec{Y}_2 &\equiv& 
G \FTpmt \eta_{j_2}|a, \alpha\ra 
\quad\mbox{or}
\\
\vec{Y}_2(\vxtp,t;t_S) &=& 
\Svxt  G(\xxtp;\xxt) \FTpmt 
\eta_{j_2}(\vxt) |a, \alpha\ra 
\eea
\bea
\vec{Z}_2 &\equiv& \gf G^{\dagger} \gf \eta_{j_2}|a, \alpha\ra 
\quad\mbox{or}
\\
\vec{Z}_2(\vxtp,t;t_S) &=& \Svyt   \gf G^{\dagger}(\xxtp;\yyt) \gf   
\eta_{j_2}(\vyt) |a, \alpha\ra 
\eea
where $\dagger$ or Hermite conjugate includes
the color, Dirac {\bf and} site indices.

Indices of these vectors are 

\be
Y^{b}_{\beta}(\vec{x}',t;t_S) = 
\sum_{\vec{x}} \sum_{c,\gamma} 
G^{b,c}_{\beta,\gamma}(\vec{x}',t;\vec{x},t_S)\xi(\vec{x})
|a,\alpha\ra_{c,\gamma}
\ee

\be
Z^{b}_{\beta}(\vec{x}',t;t_S) = 
\sum_{\vec{x}} \sum_{c,\gamma} 
\left(
G^{c,b}_{\gamma,\beta}(\vec{x},t_S;\vec{x}',t)
\right)^{*}
\xi(\vec{x}) |a,\alpha\ra_{c,\gamma}
\ee

\be
\vec{Z}^{\dagger} \cdot \vec{Y}
= \sum_{b} \sum_{\beta} \sum_{\vec{x}'}
\left( Z^{b}_{\beta}(\vec{x}',t;t_S) \right)^{*}
Y^{b}_{\beta}(\vec{x}',t;t_S) 
\ee

\subsection{Diagram H}

\bea
&H&(\vpn,-\vpn,\vpm,-\vpm)
\nonumber \\
&=& \Svxop \Svxtp \Svxo \Svxt \FTpno \FTpnt \FTpmo \FTpmt
\nonumber \\
&\times& H((\xxop),(\xxtp),(\xxo),(\xxt))
\nonumber \\
&=& \Svxop \Svxtp \Svxo \Svxt \FTpno \FTpnt \FTpmo \FTpmt
\nonumber \\
&\times& \Tr\left(
\gf \Gs(\xxo;\xxop) \gf G(\xxop;\xxtp) \gf
G(\xxtp;\xxt) \gf G(\xxt;\xxo) \right)
\nonumber \\
&=& \Sj \Svyo \Svxop \Svxtp \Svxo \Svxt \FTpno \FTpnt \FTpmo \FTpmt
\nonumber \\
&\times& \Tr\left(
\xi_{j}^{\dagger}(\vyo) \gf \Gs(\yyo;\xxop) \gf G(\xxop;\xxtp) 
\right.
\nonumber \\
&\times&
\left.
\gf G(\xxtp;\xxt) \gf G(\xxt;\xxo) \xi_{j}(\vxo) \right)
\eea
We can write a trace term as
\bea
\Tr\left(
\gf \Gs(\xxo;\xxop) \gf G(\xxop;\xxtp) \gf
G(\xxtp;\xxt) \gf G(\xxt;\xxo) \right)
\nonumber \\
= 
\sum_{a,\alpha} \la a,\alpha|
\gf \Gs(\xxo;\xxop) \gf G(\xxop;\xxtp) \gf
G(\xxtp;\xxt) \gf G(\xxt;\xxo)
|a,\alpha\ra
\eea

Then
\be
H(\vpn,-\vpn,\vpm,-\vpm;t,t_S)
= \Sj \Svxtp
\vec{Z}^{\dagger}(\vxtp,t;t_S) \FTpnt \vec{Y}(\vxtp,t;t_S)
\ee
where
\bea
\vec{Y} &\equiv& 
\Svxt \Svxo
G \FTpmt \gf G \FTpmo \xi_{j}
|a,\alpha\ra \quad \mbox{or}
\\
\vec{Y} (\vxtp,t;t_S) &=& 
\Svxt \Svxo
G(\xxtp;\xxt) \FTpmt \gf G(\xxt;\xxo) \FTpmo 
\nonumber \\
&\times& \xi_{j}(\vxo) |a,\alpha\ra
\nonumber \\
&=&
\Svxt 
G(\xxtp;\xxt) \gf \FTpmt 
\vec{Y}_1(\xxt;t_S)
\eea
Here
\be
\vec{Y}_1(\xxt;t_S)
\equiv
\Svxo
G(\xxt;\xxo) \FTpmo 
\xi_{j}(\vxo) |a,\alpha\ra
\ee
\bea
\vec{Z} &\equiv& 
\Svxop
\gf G^{\dagger} \AdjFTpno \gf \Gs^{\dagger} \gf \xi_{j}
|a,\alpha\ra \quad \mbox{or} 
\\
\vec{Z} (\vxtp,t;t_S)&=& 
\Svxop
\gf G^{\dagger}(\xxtp;\xxop) \AdjFTpno 
\Svyo \gf \Gs^{\dagger}(\xxop;\yyo) \gf 
\\
&\times& \xi_{j}(\vyo) |a,\alpha\ra
\\
&=&
\Svxop
\gf G^{\dagger}(\xxtp;\xxop) \AdjFTpno 
\vec{Z}_1(\xxop;t_S)
\eea
Here
\be
\vec{Z}_1(\xxop;t_S)  \equiv
\Svyo \gf \Gs^{\dagger}(\xxop;\yyo) \gf 
\xi_{j}(\vyo) |a,\alpha\ra
\ee

\subsection{Diagram X}

\bea
&X&(\vpn,-\vpn,\vpm,-\vpm)
\nonumber \\
&=& \Svxop \Svxtp \Svxo \Svxt \FTpno \FTpnt \FTpmo \FTpmt
\nonumber \\
&\times& X((\xxop),(\xxtp),(\xxo),(\xxt))
\nonumber \\
&=& \Svxop \Svxtp \Svxo \Svxt \FTpno \FTpnt \FTpmo \FTpmt
\nonumber \\
&\times&
\Tr\left(
\gf \Gs(\xxo;\xxop) \gf 
G(\xxop;\xxt) \gf G(\xxt;\xxtp) \gf 
G(\xxtp;\xxo) 
\right)
\nonumber \\
&=& \Sj
\Svyo \Svxop \Svxtp \Svxo \Svxt \FTpno \FTpnt \FTpmo \FTpmt
\nonumber \\
&\times&
\Tr\left(
\xi_{j}^{\dagger}(\vyo) \gf \Gs(\yyo;\xxop) \gf
G(\xxop;\xxt) \gf G(\xxt;\xxtp) 
\right.
\nonumber \\
&\times&
\left.
\gf G(\xxtp;\xxo) \xi_{j}(\vxo)
\right)
\eea
We can write a trace term as
\bea
\Tr\left(
\gf \Gs(\xxo;\xxop) \gf 
G(\xxop;\xxt) \gf G(\xxt;\xxtp) \gf 
G(\xxtp;\xxo) \right)
\nonumber \\
= 
\sum_{a,\alpha} \la a,\alpha|
 \gf \Gs(\xxo;\xxop) \gf 
G(\xxop;\xxt) \gf G(\xxt;\xxtp) \gf 
G(\xxtp;\xxo) 
 |a,\alpha\ra
\eea

Thus
\be
X (\vpn,-\vpn,\vpm,-\vpm)
= \Sj \sum_{a,\alpha} \Svxtp 
\vec{Z}(\vxtp,t;t_S)^{\dagger} \FTpnt \vec{Y}_1(\vxtp,t;t_S)
\ee
Here
\bea
\vec{Y}_1(\vxtp,t;t_S)
&\equiv& 
\Svxo G(\xxtp;\xxo) \FTpmo \xi_{j}(\vxo) |a,\alpha\ra
\\
\vec{Z}(\vxtp,t;t_S)
 &\equiv& 
\Svxt \Svxop \Svyo 
\gf G^{\dagger}(\xxtp;\xxt) \AdjFTpmt \gf G^{\dagger}(\xxt;\xxop) 
\nonumber \\
&\times&
\AdjFTpno \gf \Gs^{\dagger}(\xxop;\yyo) \gf \xi_{j}(\vyo) 
|a, \alpha\ra
\nonumber \\
&=& \Svxt
\gf G^{\dagger}(\xxtp;\xxt) 
\AdjFTpmt 
\vec{Z}'(\vxt,t_S;t)
\\
\vec{Z}'(\vxt,t_S;t)
&\equiv& 
\Svxop \Svyo
\gf G^{\dagger}(\xxt;\xxop) 
\AdjFTpno \gf \Gs^{\dagger}(\xxop;\yyo) \gf 
\nonumber \\
&\times& \xi_{j}(\vyo) 
|a,\alpha\ra
\nonumber \\
&=&
\Svxop
\gf G^{\dagger}(\xxt;\xxop) 
\AdjFTpno 
\vec{Z}_1(\vxop,t;t_S)
\nonumber \\
\vec{Z}_1(\vxop,t;t_S) 
&\equiv& 
\Svyo
\gf \Gs^{\dagger}(\xxop;\yyo) \gf \xi_{j}(\vyo) 
|a,\alpha\ra
\eea

\newpage

\end{document}